\documentclass[aps,prd,amssymb,nobibnotes, amsmath,twocolumn,nofootinbib,superscriptaddress,floatfix,preprintnumbers]{revtex4-1}

\newcommand{\code}{{\texttt{MUNHECA}}}
\usepackage{graphicx,color}
\usepackage{dcolumn}
\usepackage{bm}
\usepackage{hyperref}
\usepackage{amsmath}
\usepackage{mathtools}
\usepackage{multirow}
\usepackage{stackrel}
\usepackage[lofdepth,lotdepth,caption=false]{subfig}
\usepackage{tikz-feynman}
\usepackage{comment}
\usepackage{xcolor}
\def\d{{\rm d}}

\begin{document}

\begin{flushleft}
LAPTH-053/23
\end{flushleft}

\title{Neutrinos from muon-rich ultra high energy electromagnetic cascades:\\ The \texttt{MUNHECA} code}

\author{AmirFarzan Esmaeili}
\email{a.farzan.1993@aluno.puc-rio.br}
\affiliation{Departamento de Física, Pontifícia Universidade Católica do Rio de Janeiro, Rio de Janeiro 22452-970, Brazil}

\author{Arman Esmaili}
\email{arman@puc-rio.br}
\affiliation{Departamento de Física, Pontifícia Universidade Católica do Rio de Janeiro, Rio de Janeiro 22452-970, Brazil}

\author{Pasquale Dario~Serpico}
\email{serpico@lapth.cnrs.fr}
\affiliation{LAPTh, CNRS, USMB, F-74940 Annecy, France}

\begin{abstract}

An ultra high energy electromagnetic cascade, a purely leptonic process and initiated by either photons or $e^\pm$, can be a source of high energy neutrinos. We present a public \texttt{python3} code, \code{}, to compute the neutrino spectrum by taking into account various QED processes, with the cascade developing either along the propagation in the cosmic microwave background in the high-redshift universe or in a predefined photon background surrounding the astrophysical source. The user can adjust various settings of \code{}, including the spectrum of injected high energy photons, the background photon field and the QED processes governing the cascade evolution. We improve the modeling of several processes, provide examples of the execution of \code{} and compare it with some earlier and more simplified estimates of the neutrino spectrum from electromagnetic cascades.

\end{abstract}

\maketitle

\section{Introduction\label{sec:intro}}

Unveiling the still unknown ultra high energy astrophysical sources in the sky greatly benefits of a multimessenger approach, combining information from charged cosmic rays with the one carried by photons and neutrinos. In turn, implementing this program requires mastering the microphysics processes shaping the spectra and the energy repartition in the respective species. In this context, we have recently studied a purely leptonic channel for production of ultra high energy neutrinos, linked to muon-producing QED processes and most notably muon pair production (MPP) $\gamma\gamma\to \mu^+\mu^-$~\cite{Esmaeili:2022cpz}. This project required revisiting the dynamics of ultra high energy electromagnetic cascades\footnote{By ``ultra high energy cascade'' we specifically refer here to cascades with the initial center of mass energy above the MPP threshold, i.e. Mandelstam $s$-variable $s\gtrsim 4m_\mu^2$, where $m_\mu$ is the muon mass. For the case of cascades from pristine sources developing during propagation over cosmological distances onto the CMB photon target, this mostly affects observable energies at the Earth above the ones currently observed in neutrino telescopes, i.e. $E\gtrsim 10^{16}\div 10^{17}~{\rm eV}$ (see Sec.~\ref{sec:example} for details). The astrophysical sources that are emitting particles in this energy range are called ``ultra high energy sources'' in this article. For electromagnetic cascades inside the astrophysical sources considered in section~\ref{sec:example}, it is straightforward to derive the energetic regime of interest in the Lab  by substituting the CMB with the chosen background photon field.}. It turns out that the rarely considered double pair production (DPP) $\gamma\gamma\to e^+e^-e^+e^-$ plays an important role for a quantitative assessment of the relevance of the MPP: It acts as a competitor energy-loss process in the regime where the frequent alternating electron pair production (EPP) $\gamma\gamma\to e^+e^-$ and inverse Compton scattering (ICS) $\gamma\, e\to \gamma\,e$ processes (in the deep Klein-Nishina limit) are rather ineffective in degrading the photon energy. Also, other processes such as electron triplet production (ETP)\footnote{In the literature this process is sometimes simply called triplet pair production ({\it e.g.}, see~\cite{Lee:1996fp}). However, we prefer to use the more self-explanatory name ``electron triplet production" in this article.} $e\gamma\to ee^-e^+$, electron muon-pair production (EMPP) $e\gamma\to e\mu^-\mu^+$ and charged pion pair production $\gamma\gamma\to\pi^+\pi^-$ can affect, typically at the (1-10)\% level, the electromagnetic cascade development at ultra high energies, especially the emerging neutrino spectrum. For ease of reference, a list of acronyms for the processes considered in this article is given in Table~\ref{tab:acr}. 

\begin{table}[h!]
\caption{List of the processes considered in this article and their acronyms.\label{tab:acr}}
    \tiny
    \centering
    \begin{tabular}{|c|c|c|}
    \hline
    Process  & Name & Acronym \\
    \hline
    $\gamma\gamma\to e^+e^-$ & Electron Pair Production & EPP\\
    \hline
    $\gamma\gamma\to \mu^+\mu^-$ & Muon Pair Production & MPP\\
    \hline
    $\gamma\gamma\to e^+e^-e^+e^-$ & Double Pair Production & DPP\\
    \hline
    $\gamma\gamma\to \pi^+\pi^-$ & Charged Pion Pair Production & CPPP\\
    \hline
    $e\gamma\to e\gamma$ & Inverse Compton Scattering & ICS\\
    \hline
    $e\gamma\to e\mu^+\mu^-$ & Electron Muon-Pair Production & EMPP\\
    \hline
    $e\gamma\to ee^+e^-$ & Electron Triplet Production & ETP\\
    \hline
\end{tabular}
\end{table}

In Ref.~\cite{Esmaeili:2022cpz} the electromagnetic cascade development has been studied for the propagation of ultra high energy photons emitted at cosmological distances (redshifts $z\sim$~5-15) interacting on Cosmic Microwave Background (CMB) target photons. This is basically the only relevant target expected in the high-redshift Universe. However, provided that the magnetization of the environment is low enough, the same microphysics can play a role in the processing of moderately high energy photons, say of energy $\sim\mathcal{O}(100)$~TeV, within an astrophysical source where they interact with thermal X-ray environmental photons. The neutrino flux emerging from the electromagnetic cascade development inside the source serves at least in principle as a counterexample to the common consensus that neutrino detection is the smoking gun signal for hadronic processes in the source.                

Motivated by these remarks, {\it i.e.} the possibility of neutrino production in a purely leptonic framework, and aware of the complication brought by several relevant processes in the development of the electromagnetic cascade, we developed a public code, \code\footnote{\textbf{MU}ons and \textbf{N}eutrinos in \textbf{H}igh-energy \textbf{E}lectromagnetic \textbf{CA}scades. The code can be downloaded from \url{https://github.com/afesmaeili/MUNHECA.git}}, facilitating the computation of the emergent neutrino spectrum. For the processes of interest, a detailed calculation of the cross sections and inelasticities are either missing or rarely found in the literature. For example, the only relatively recent (still, 15 years old!) reference we could find dedicated to the DPP is Ref.~\cite{Demidov:2008az}, in turn building up on much older cursory studies~\cite{Cheng:1970zb,Brown:1973onk} used till then (see e.g.~\cite{Lee:1996fp}). In~\cite{Esmaeili:2022cpz}, we relied on some approximations suggested by the results of ref.~\cite{Demidov:2008az} for a first treatment. Here we recompute these processes with the goal to perform dedicated studies aimed at checking existing estimates and, when needed, improving them, by assessing the impact of these processes on the cascade development.  The most widely used cascade simulation code, included in~\texttt{CRPropa3.2},  only  accounts for DPP and ETP, but does not include any leptonic neutrino-generating processes i.e. MPP, CPPP and EMPP~\cite{crpropa:2022lkd}. Therefore, it yields no neutrino spectrum from a purely leptonic cascade development. Moreover, the DPP in this existing framework~\cite{eleca:2018asl} is based on the approximations of ref. \cite{Demidov:2008az}.  

A qualitative description of the electromagnetic cascade development at ultra high energies and the impact of these processes is given in section~\ref{sec:cas} while the details of the evaluation of cross sections, differential cross sections and inelasticities of specific processes are reported in the appendices. 
\code{} computes the neutrino spectrum produced during the propagation of high energy photons either over cosmological distances or within the astrophysical sources. The user-friendly structure of the code allows the user to choose between various high energy photon injection spectra, target photon spectra and the processes to be included in the development of electromagnetic cascade. The structure of the code and its features are described in section~\ref{sec:code}. In section~\ref{sec:example} we report the output of the code for two case studies: i) The propagation of monochromatic ultra high energy photons injected at high redshift in the cosmological background, which can be directly compared with our previous approximate results in~\cite{Esmaeili:2022cpz}. ii) The case of cascade development inside the source, a proxy for the recently identified neutrino source NGC 1068~\cite{IceCube:2022der}, which can be compared with the more simplistic estimate presented in~\cite{DanHooper:2023dls}. Finally, in section~\ref{sec:conc} we discuss our results and report our conclusions and perspectives.

\section{Electromagnetic cascades at ultra high energies\label{sec:cas}}

The relevance of electromagnetic cascade phenomena was remarked soon after the discovery of the CMB and has been scrutinized in the seminal paper~\cite{Berezinsky:1975zz} as a handle to probe cosmogenic neutrinos. The conventional development of the cascade at energies above the EPP threshold is rooted in the features of EPP and ICS at high energies, mainly their large inelasticities\footnote{See Appendix~\ref{append:inelasticity} for a review of the inelasticity concept and its characteristics.}. In both the EPP and ICS, almost all the energy of the initial high energy particle (one of the photons in EPP and the electron or positron in ICS) will be transferred to one of the produced particles (so-called leading particle) which is one of the $e^+$ or $e^-$ in EPP and the photon in ICS. This regeneration of the leading particle in each step in EPP and ICS makes the energy degradation rather slow and quasi-continuous. 

\begin{figure}[t!]
    \centering
    \includegraphics[width=0.48\textwidth]{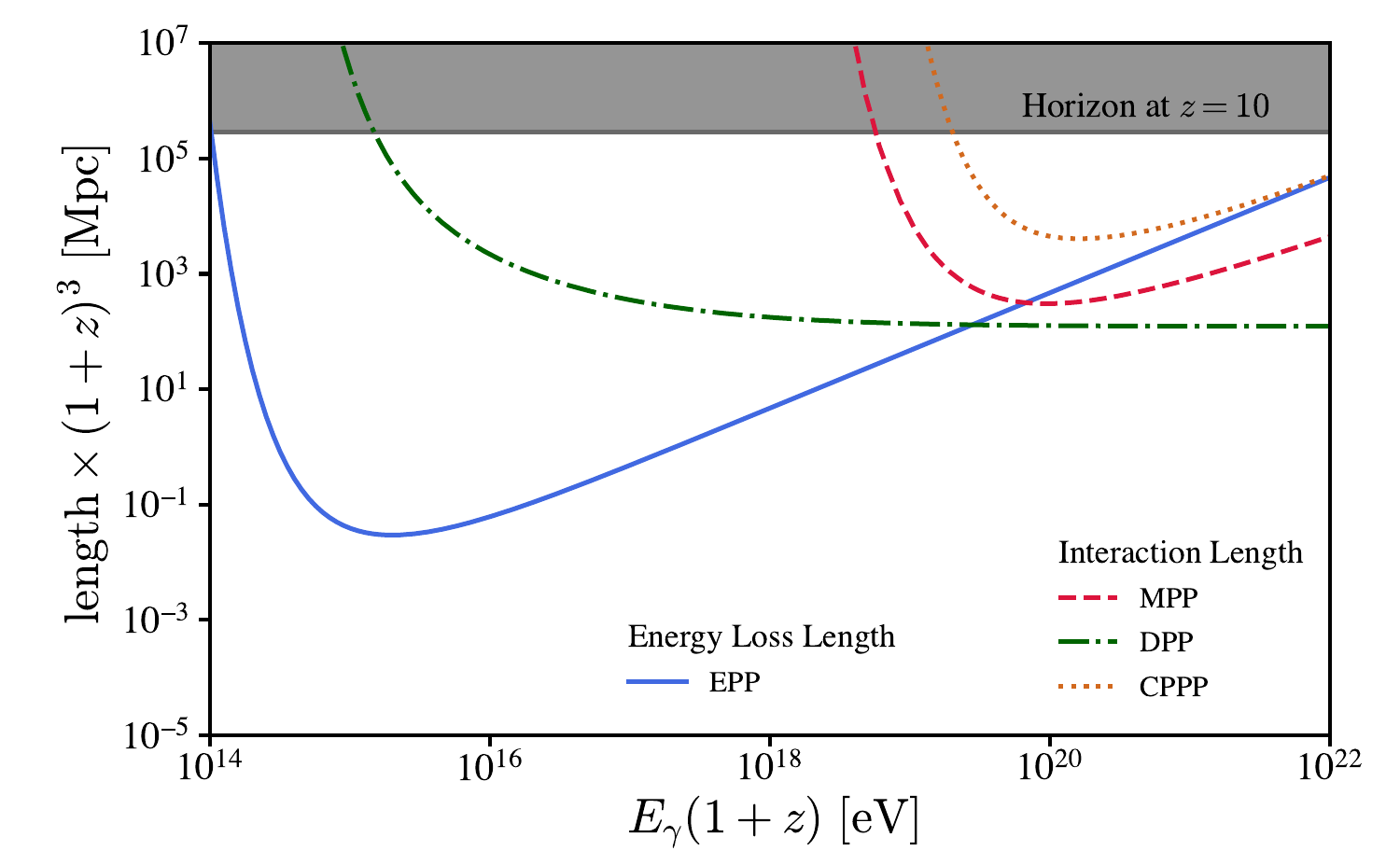}
    \caption{Characteristic lengths of photon-photon processes considered in this work, if interacting with CMB target photons. The EPP energy loss length is compared with the MPP, DPP and CPPP interaction lengths. The Hubble horizon at $z=10$ is shown by the gray line.}
    \label{fig:gg_lengths}
\end{figure}

However, once raising the energy scale above the appropriate threshold, the feasibility of muon production changes the energy repartition in the cascade development. The muon production, primarily through MPP, is relevant in spite of its cross section being smaller than the EPP one, since the energy loss length of electrons is smaller than the interaction length of muon production processes. For example, for the case of cascade development on the CMB, figure~\ref{fig:gg_lengths} shows the interaction length of MPP (dashed red curve) and the energy loss length of electrons via EPP (solid blue curve), where the predominance of muon production at $E_\gamma(1+z)\gtrsim10^{20}$~eV is evident ($z$ being the injection redshift of the high energy photon with energy $E_\gamma$). Notice that the vertical and horizontal axes in figure~\ref{fig:gg_lengths} are scaled by $(1+z)^3$ and $(1+z)$, respectively, making the curves valid for any injection redshift. However, whenever the MPP interaction length is smaller that the EPP energy loss length, the DPP interaction length plays the leading role in photon-photon interactions, as illustrated by the dot-dashed green curve in figure~\ref{fig:gg_lengths}. The details of the DPP process are reported in Appendix~\ref{append:dpp}, though for the qualitative discussion which is the goal of the rest of this section we just need some notions on its inelasticity: At high energies, one of the produced pairs of $e^\pm$ takes almost all of the initial energy and shares it between the $e^+$ and $e^-$ roughly in a ratio $1:3$. As a result, the impact of the DPP on the cascade development is twofold: First, it effectively reduces the energy of the initial photon by a factor $\sim2$, contrary to the energy degradation via EPP which is gradual. Second, it doubles the number of $e^\pm$ in the cascade, increasing the multiplicity of MPP events and hence the neutrino yield (as long as one is above threshold). The two effects compete against each other, especially at $E_\gamma(1+z)\gtrsim10^{21}$~eV where the DPP suppresses the muon (and neutrino) production while the increase in multiplicity of MPP occurrence significantly increases the neutrino yield at $E_\gamma(1+z)\simeq 10^{20}$~eV. A minor but still appreciable effect is due to CPPP (whose interaction length is depicted by the dotted orange curve in figure~\ref{fig:gg_lengths}) which further raises the neutrino yield at $E_\gamma(1+z)\gtrsim10^{19}$~eV. For the cross section of CPPP and its inelasticity we have used the Born approximation as in~\cite{Brodsky:1971ud}, treating the charged pions as pointlike scalar particles. Within the same approximation, the cross section for pairs of neutral pions vanishes. Although the Born approximation only provides a rough description of a more complex process (with hadronic resonances eventually playing a role), given the sub-leading contribution of the pion channel, we deemed it sufficient here.

The horizontal line in figure~\ref{fig:gg_lengths} shows the Hubble horizon of universe at $z=10$ laying above the curves which justifies our omission of cosmological evolution effects in the cascade development. We used $H_0=67.4~{\rm km~s^{-1}~Mpc^{-1}}$, $\Omega_M=0.315$ and $\Omega_{\Lambda}=0.685$, according to \cite{planck:2018omg}. For all practical purposes, the cascade happens instantaneously at the injection redshift, which only determines the density and energy of the target photons~\cite{Esmaeili:2022cpz}.  We estimate the error of this approximation to be $\Delta z\sim\mathcal{O}(10^{-4})$.

\begin{figure}[h!]
    \centering
    \includegraphics[width=0.48\textwidth]{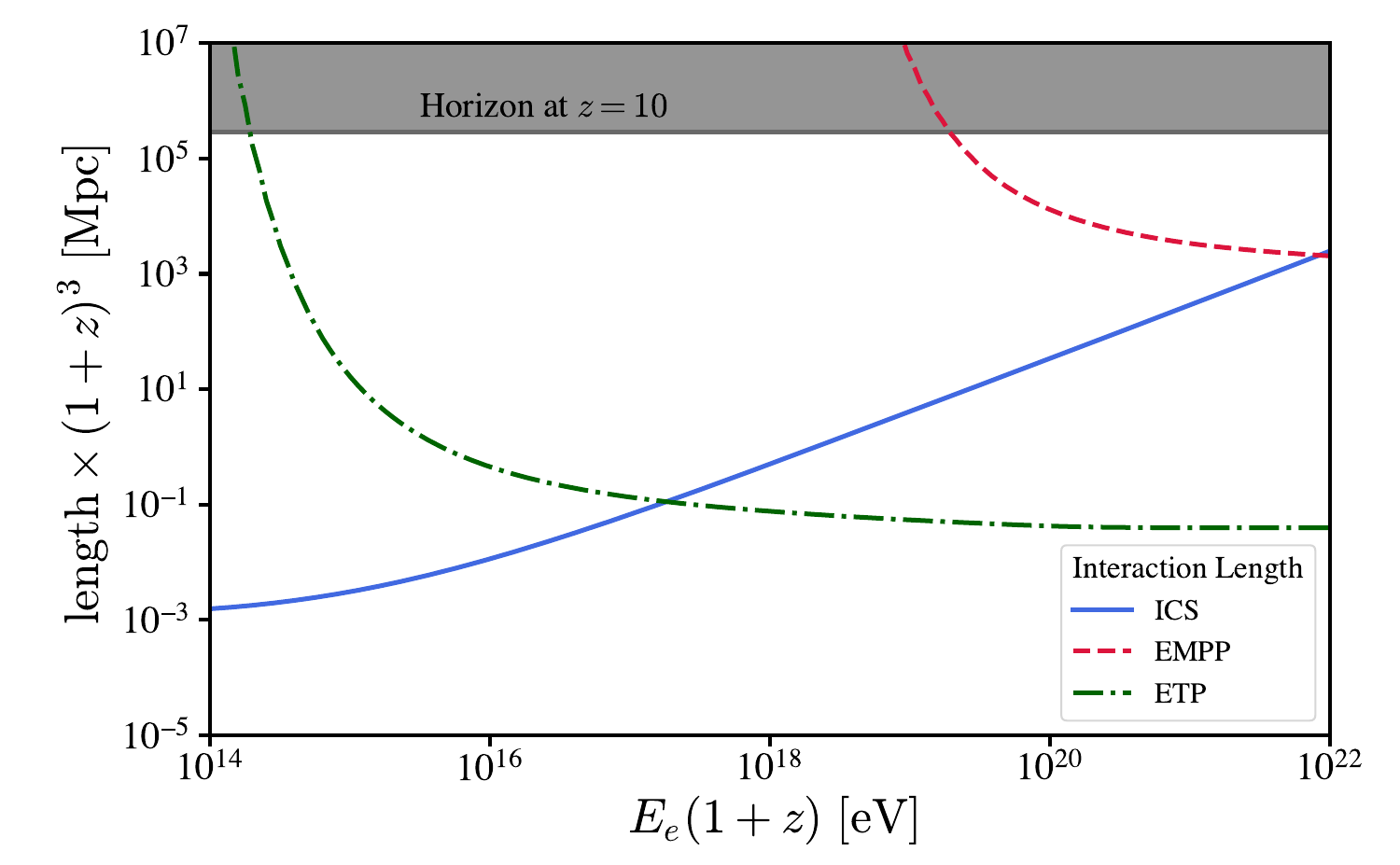}
    \caption{Interaction length of the electron-photon processes considered in this work, if interacting with CMB target photons.}
    \label{fig:eg_lengths}
\end{figure}

The characteristic lengths for electron-photon interaction processes are shown in figure~\ref{fig:eg_lengths}. At low energies we recover the standard picture where the ICS (whose interaction length is depicted by the solid blue curve) governs the cascade development. However, already at $E_\gamma(1+z)\gtrsim10^{17}$~eV the ETP becomes the most frequent process. Yet, the peculiar energy share among the three produced electrons/positrons in ETP renders it ineffectual for the cascade development: The pair of $e^\pm$ produced in ETP only carries a very small fraction ($\sim10^{-5}$) of the initial high energy interacting $e$. Two of the three particles thus drop below the MPP threshold, while the third one is basically unaffected. In summary, the ETP only causes the leading particle to lose a negligible fraction of its energy, and it can be safely ignored in our simulation (as typically done in the specialized literature). Another muon-producing process in the $e\gamma$ interaction is the EMPP, whose interaction length is shown by the dashed red curve in figure~\ref{fig:eg_lengths}. It only becomes notable at $E_\gamma(1+z)\gtrsim10^{21}$~eV. A brief report on the evaluation of its cross section and inelasticity is provided in Appendix~\ref{append:EMPP}. 

A remark on ICS is in order. The cross section of double Compton scattering (DCS) $e\gamma\to e\gamma\gamma$ logarithmically increases at high energy and, for $\epsilon_\gamma\gtrsim10^{11}$~eV, it is larger than the conventional ICS cross section~\cite{Ram:1971pi}, where $\epsilon_\gamma$ is the energy of photon in the rest frame of electron. The extra photon emitted in DCS is a soft photon and the cross section is in fact infrared-divergent, {\it i.e.} in the limit of zero energy of the soft photon. However, two other related QED processes should be considered: the \textit{multiple} Compton scattering $e\gamma\to e+n\gamma$, where $n\geq3$, and radiative corrections to the conventional (single) Compton scattering. The sum of the amplitudes for these processes not only cures the infrared divergence in cross section but also cancels the apparent increase of DCS cross section at high energy~\cite{Gould:1979pw}, such that the total cross section of these processes only differs at the few percent level from the leading-order cross section of the single Compton scattering at the energy range of our interest. All the emitted extra soft photons in the multiple Compton scattering land in the very low energy range and their impact on the evaluation of cascade evolution at ultra high energy is negligible. Hence, in our simulation we use the conventional ICS cross section, being aware that, apart for a small logarithmic factor, it accounts also for the above-mentioned effects.

The development of an electromagnetic cascade at ultra high energies is intrinsically stochastic. If focusing on the energy drainage to neutrinos, the diverse processes discussed earlier contribute unequally to the formation of the neutrino spectrum. To capture the multiple occurrences of muon (and occasionally pion) producing processes, realizable due to electron-proliferating DPP process, and to handle the abrupt energy degradation, a Monte Carlo simulation of the cascade evolution is required. This is handled by the \code{} code, described in the next section.

\section{\code: structure and features\label{sec:code}}

To compute the full-fledged cascade evolution including all the relevant $\gamma\gamma$ and $e\gamma$ interactions discussed in the previous section, we have developed a Monte Carlo code named \code. The \code, written in \texttt{python3}, can be used to compute the neutrino spectrum from electromagnetic cascade at the ultra high energies, either inside an astrophysical source characterized by arbitrary target spectrum, or in the cosmological propagation setting with the CMB as target. \code{} consists of two parts. The main part, using a Monte Carlo algorithm, tracks the particles along the cascade evolution and produces seven output files recording all the muon, pion, photon and electron energies at the end of cascade evolution, as well as the number of occurrences of neutrino-producing processes (MPP, CPPP and EMPP) for each of the $n$ (to be chosen by the user) monochromatic-energy photons injected into the Monte Carlo realization. The secondary piece of code reads instead the muon and pion outputs of the main part and yields the corresponding neutrino spectra at the Earth. 

\subsection{Installation and Execution}\label{subsec:install}

\code{} uses the \texttt{numpy} and \texttt{Scipy} packages and does not need any additional installation. To run the code, a \texttt{.txt} file should be created which contains the input to the code. A self-explanatory example of the input file, the \texttt{test.txt}, is provided in the \texttt{/Work} directory. Further description of the format and some clarifications are provided in section.~\ref{subsec:inout}.   

To run the Monte Carlo session, the following command can be used \\

\noindent\texttt{python3 [PATH]/run/Main.py} \\
\texttt{[INPUT\_PATH]/[InputFileName].txt}\footnote{The executable \texttt{Main.py} is located inside the \texttt{run} directory. The execution should be addressed to \texttt{/run} as it is indicated in the commands.}\\

After the execution, the results of the session are saved, together with a copy of the input file, into the \texttt{/results} directory. The copied input file name, depending on the chosen injection spectrum (\texttt{Monochrome} or \texttt{PowerLaw}) are named \texttt{0M\_input.txt} or \texttt{0S\_input.txt}, respectively.

To obtain the neutrino fluxes at the Earth, the relevant syntax is\\

\noindent\texttt{python3 [PATH]/run/nuSpec.py} \\ \texttt{[PATH]/results/[DESTINATION\_DIR]/0M\_input.txt} \\

\noindent or \\

\noindent\texttt{python3 [PATH]/run/nuSpec\_Weight.py} \\  \texttt{[PATH]/results/[DESTINATION\_DIR]/0S\_input.txt } \\

\noindent for \texttt{Monochrome} or \texttt{PowerLaw} injection, respectively. 

These commands result in the creation of the file \texttt{NEUTRINO\_EARTH.txt} in the same destination directory, which contains a table of the neutrinos fluxes at the Earth, including all-flavors and each flavor separately.

\subsection{Inputs and Outputs}\label{subsec:inout}

As we mentioned above, the input file format is important and discrepancies may cause \texttt{ValueErrors}. Therefore, in this section we describe the structure and the input keywords based on the provided template \texttt{test.txt}.

Before describing the input structure, let us make a remark. In the input file, the letter capitalization and the place of the colons and the spaces are important. Any parameter inside the input file should be followed by a \textit{colon} (:) and a \textit{space} as is written in this section. 

The input file contains four sets of parameters: The choice of processes in the cascade evolution, the background photon field choice, the injection settings and the output file names and directory. Among all the six implemented interactions (5 leptonic + 1 hadronic) the EPP and ICS are activated by default, while the user can turn on or off the MPP, DPP, EMPP and CPPP; for example by selecting \texttt{MPP: ON} or \texttt{MPP: OFF}; equivalent syntax applies to \texttt{DPP}, \texttt{EMPP} and \texttt{CPPP}.

The background photon field (\texttt{Source}) can be chosen among: i) CMB (in case of propagation in a cosmological setting), ii) Black-body, and iii) Power-law (in case of propagation inside a source) distributions by setting \texttt{Source: CMB}, \texttt{Source: BlackBody} or \texttt{Source: PowerLaw}, respectively. Once the background photon field is chosen, its characteristics should be set. For the CMB the injection redshift, for the black-body its temperature (in eV) and for the power-law the energy index are the parameters that should be written in front of the \texttt{Redshift}, \texttt{BB\_temp} and \texttt{PL\_index}, respectively. Also, for the \texttt{BlackBody} and \texttt{PowerLaw} photon fields, the energy range of the spectrum should be set by using the keywords \texttt{BB\_Emin}, \texttt{BB\_Emax}, \texttt{PL\_Emin} and \texttt{PL\_Emax}.

The photon injection setting can be chosen between \texttt{INJ\_Spectrum: Monochrome} and \texttt{INJ\_Spectrum: PowerLaw}. The relevant input parameters for the monochromatic injection, \texttt{Monochrome}, are the number of Monte Carlo realizations, \texttt{Number\_of\_photons}, and the injected photon energy, \texttt{E\_gamma}. For the \texttt{PowerLaw} injection there are nine parameters: the \texttt{INJ\_SPEC\_Index} fixes the energy index $\alpha$; the energy range of injection $[E_{\rm min},E_{\rm max}]$ is defined by \texttt{INJ\_Emin} and \texttt{INJ\_Emax} and characterized by sharp cutoffs. Alternatively,  the minimum and maximum energies can be characterized by exponential cutoff scales $E_0$ and $E^\prime_0$, respectively, controlled by \texttt{ExpCut\_LowEnergy} and \texttt{ExpCut\_HighEnergy}. In this case, the injected spectrum takes the form
$$\frac{{\rm d}N}{{\rm d}E} \propto E^{-\alpha} e^{-E/E^\prime_0}e^{-E_0/E}~.$$
All the energies should be provided in eV. To implement the low and high energy exponential cutoffs, respectively the \texttt{EXP\_LOW\_CUTOFF} and \texttt{EXP\_HIGH\_CUTOFF} should be set to \texttt{ON}. The \texttt{NUMBER\_OF\_BINS} and \texttt{PHOTON\_PER\_BIN} set the number of bins in the energy range and the number of photons injected in each bin, respectively. The \texttt{PowerLaw} injection is basically an automatized example of a user-defined arbitrary injection spectrum which will be described in the next section. 

There are two other parameters common to both spectra: the \texttt{BREAK\_Energy} is the energy at which the simulation stops and the \texttt{Redshift} which defines the redshift of the source. As we mentioned earlier, only if the \texttt{CMB} background photon field is chosen, the redshift will be needed in the simulation. However, for any choice of the background photon field, the source redshift is still needed for the computation of neutrino flux at the Earth. The break point of the Monte Carlo, in principle, can be set to any reasonable value, even below the EPP threshold and of course this will give the complete cascade evolution down to this energy. However, for the purpose of computing the neutrino spectrum generated during the cascade evolution, a break energy just below the MPP threshold saves a considerable computational time. If interested in lower energies, it is more effective to re-inject the outputs of the evolution till the MPP threshold into a new run with just the EPP, DPP and ICS switched \texttt{ON}.

At the end of the \texttt{test.txt} file, there are eight lines controlling the destination directory and output files names:
\begin{itemize}
    \item \texttt{DESTINATION\_DIR}: is the name of the directory in which the outputs will be written. This directory will be created inside the \texttt{/results} as soon as the code starts running. 
    \item \texttt{Muon\_Output} and \texttt{Pion\_Output}: contain, respectively, all the muon and charged pion energies produced during the cascade evolution.
    \item \texttt{Gamma\_Output} and \texttt{Electron\_Output}: are the histogram list of the photons and electrons below the break energy, each file consisting of three columns. The first column is the center of each bin, the bin widths are written in the second column and the third contains the number of photons/electrons in the bin. To reduce the RAM consumption and the running time, a histogram of the electrons and photons with a fine binning is performed after each realization.
    \item \texttt{MPP\_Output}, \texttt{EMPP\_Output} and \texttt{CPPP\_Output}: are the occurrences of MPP, EMPP and CPPP interactions, respectively, in each Monte Carlo realization. The size of the lists in these files is equal to the number of injected photons.    
\end{itemize}

\subsection{Neutrino Spectrum\label{sec:nu}}

The output files of the simulation together with a copy of the input file are written in the \texttt{/results/[DESTINATION\_DIR]} directory, as we mentioned in section~\ref{subsec:install}. After executing the neutrino flux computation command, the \texttt{nuSpec} and \texttt{nuSpec\_Weight} will be fed by the copied input files and the muons and pions tables. The neutrino fluxes (all-flavor and the $\nu_e$, $\nu_\mu$ and $\nu_\tau$ flavors separately) at the Earth are computed by using the neutrino spectra from muon and pion decays (see appendix A of~\cite{Esmaeili:2022cpz}) and taking into account neutrino mixing parameters fixed to the best fit parameters from \cite{nufit:2022asd} and \cite{esteban:2020nus}.

The neutrino spectra files provide the sum of the neutrino and anti-neutrino fluxes produced in the course of cascade evolution. Due to the matter-antimatter symmetry of the processes, the individual neutrino or anti-neutrino flux is simply half of the total flux.

The module \texttt{nuSpec\_Weight} performs the weighting of the neutrino flux according to the injected high energy photon spectrum. The weights, normalizing the neutrinos spectrum in accordance with the number of photons injected in each bin of injected photon spectrum, are defined by  
\begin{equation}\label{eq:weight}
    w_i = \frac{\int_{E_{i-1}}^{E_i}f(E)\d E}{\int_{E_{\rm min}}^{E_{\rm max}}f(E)\d E}~,
\end{equation}
where $f(E)$ is the injected photon spectrum, $E_i$ and $E_{i-1}$ are the bin edges of the $i$-th bin, $E_{\rm min}$ = \texttt{INJ\_Emin} and $E_{\rm max}$ = \texttt{INJ\_Emax}. For the power-law spectrum
$$f(E) = E^{-\alpha}\exp(-k_{\rm hc}E/E_{\rm hc})\exp(-k_{\rm lc}E_{\rm lc}/E)~,$$
where $\alpha$ = \texttt{INJ\_SPEC\_Index}, $E_{\rm hc}$ = \texttt{ExpCut\_HighEnergy}, $E_{\rm lc}$ = \texttt{ExpCut\_LowEnergy} and $k_{\rm hc/lc} = 1~(0)$ if \texttt{EXP\_HIGH\_CUTOFF}/\texttt{EXP\_LOW\_CUTOFF} are set to \texttt{ON} (\texttt{OFF}). The power-law spectrum, with or without cutoff, is taken care of by the module automatically; instead, for a general injected spectrum $f(E)$, the user can perform the weighting by a separate script. An example of $f(E)$ different than the power-law is given in section~\ref{sec:NGC1068}.

\section{Case Studies\label{sec:example}}

In this section we illustrate the capabilities of \code{} by applying it to two cases of astrophysical interest. In section~\ref{sec:mono}, we re-assess our previous results in~\cite{Esmaeili:2022cpz} for the propagation of ultra high energy monochromatic photons over cosmological distances at high redshift and discuss the consequences of the newly included processes and refinements. Section~\ref{sec:NGC1068} is devoted to the cascade development inside an astrophysical source, re-evaluating the neutrino yields for the model proposed by~\cite{DanHooper:2023dls} for the neutrino source NGC1068 with the more complete physics incorporated in \code. 

\subsection{Monochromatic Photon Injection\label{sec:mono}}

The neutrino spectrum from the cosmological propagation of high energy monochromatic photons injected at a redshift $z$ has been calculated in~\cite{Esmaeili:2022cpz}, where only the EPP, ICS, MPP and DPP have been considered and the approximation described in~\cite{Demidov:2008az} was used for DPP. Here we re-evaluate the neutrino spectrum with \code{}, taking into account the other processes discussed in this article and the re-assessment of DPP detailed in Appendix~\ref{append:dpp}. The content of the input file of \code{} for 5000 monochromatic photons with energy $E_\gamma=10^{21}$~eV injected at $z=10$ is 

\texttt{MPP: ON}

\texttt{DPP: ON}

\texttt{EMPP: ON}

\texttt{CPPP: ON}

\texttt{Source: CMB}

\texttt{Redshift: 10}

\texttt{INJ\_Spectrum: Monochrome}

\texttt{E\_gamma: 1e+21}

\texttt{Number\_of\_photons: 5000}

\begin{figure}[t!]
    \centering
    \includegraphics[width=0.48\textwidth]{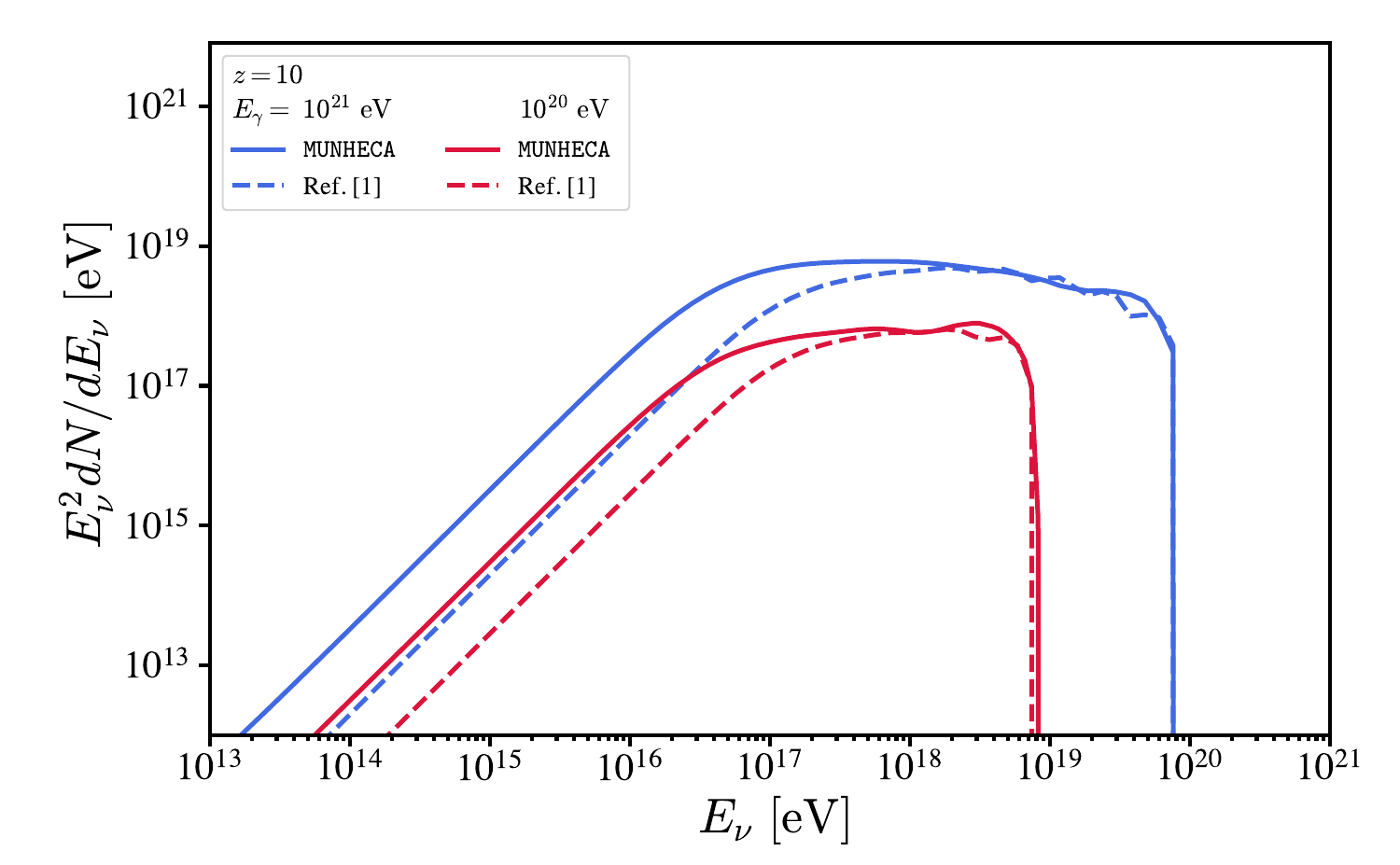}
    \caption{The neutrino spectrum from cosmological propagation of photons injected at $z=10$ with energy $E_\gamma=10^{21}$~eV (solid blue curve) and $E_\gamma=10^{20}$~eV (solid red curve) obtained by \code. For comparison, the neutrino spectra computed in~\cite{Esmaeili:2022cpz}, by considering just the EPP, ICS, MPP and DPP, are shown by the dashed curves.}
    \label{fig:mono_nuSpec}
\end{figure}

\texttt{BREAK\_Energy: 1.1e+17 } 
\\
The runtime for this input on one core of Intel(R) 12th Gen processor (i7-12700K) is $\sim 5$ hours. The produced neutrino spectrum is depicted by the solid blue curve in figure~\ref{fig:mono_nuSpec}. For comparison, the dashed blue curve depicts the neutrino spectrum from the setup of~\cite{Esmaeili:2022cpz}. The solid and dashed red curves in figure~\ref{fig:mono_nuSpec} are for $E_\gamma=10^{20}$~eV, evaluated by \code{} and in Ref.~\cite{Esmaeili:2022cpz}, respectively. We see that the results are similar down to about $10^{18}$~eV, but the more complete calculation of \code{} yields a flux increase by almost one order of magnitude at energies $\lesssim10^{18}$~eV.

\begin{figure}[t!]
    \centering
    \includegraphics[width=0.49\textwidth]{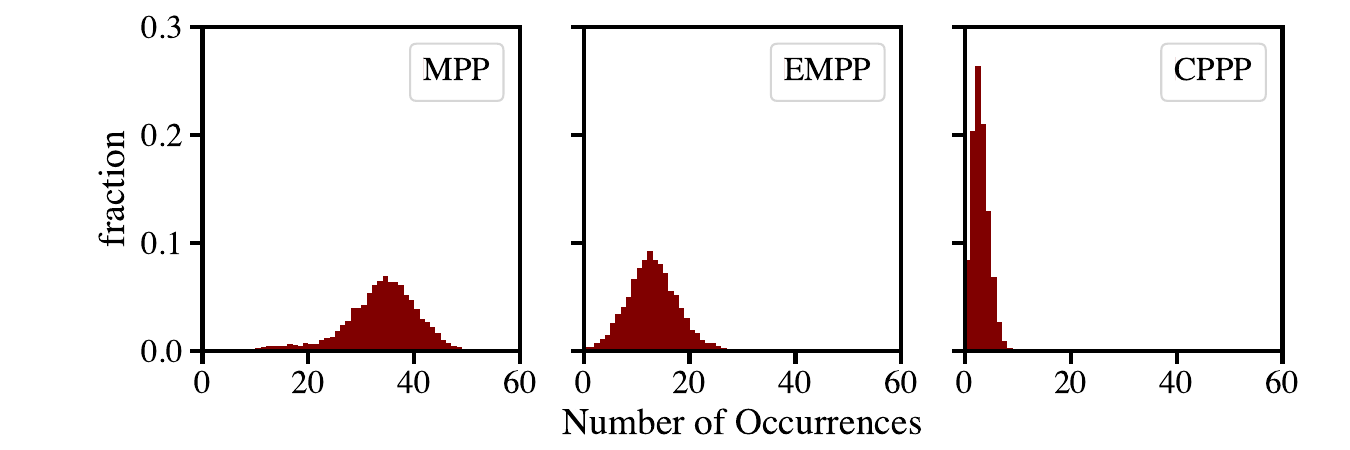}
    \caption{The distribution of the occurrences of MPP, EMPP and CPPP in $5000$ realization of photon injection at $z=10$ with energy $E_\gamma = 10^{21}$~eV.}
    \label{fig:num_of_interactions}
\end{figure}

\begin{figure}[t!]
    \centering
    \includegraphics[width=0.48\textwidth]{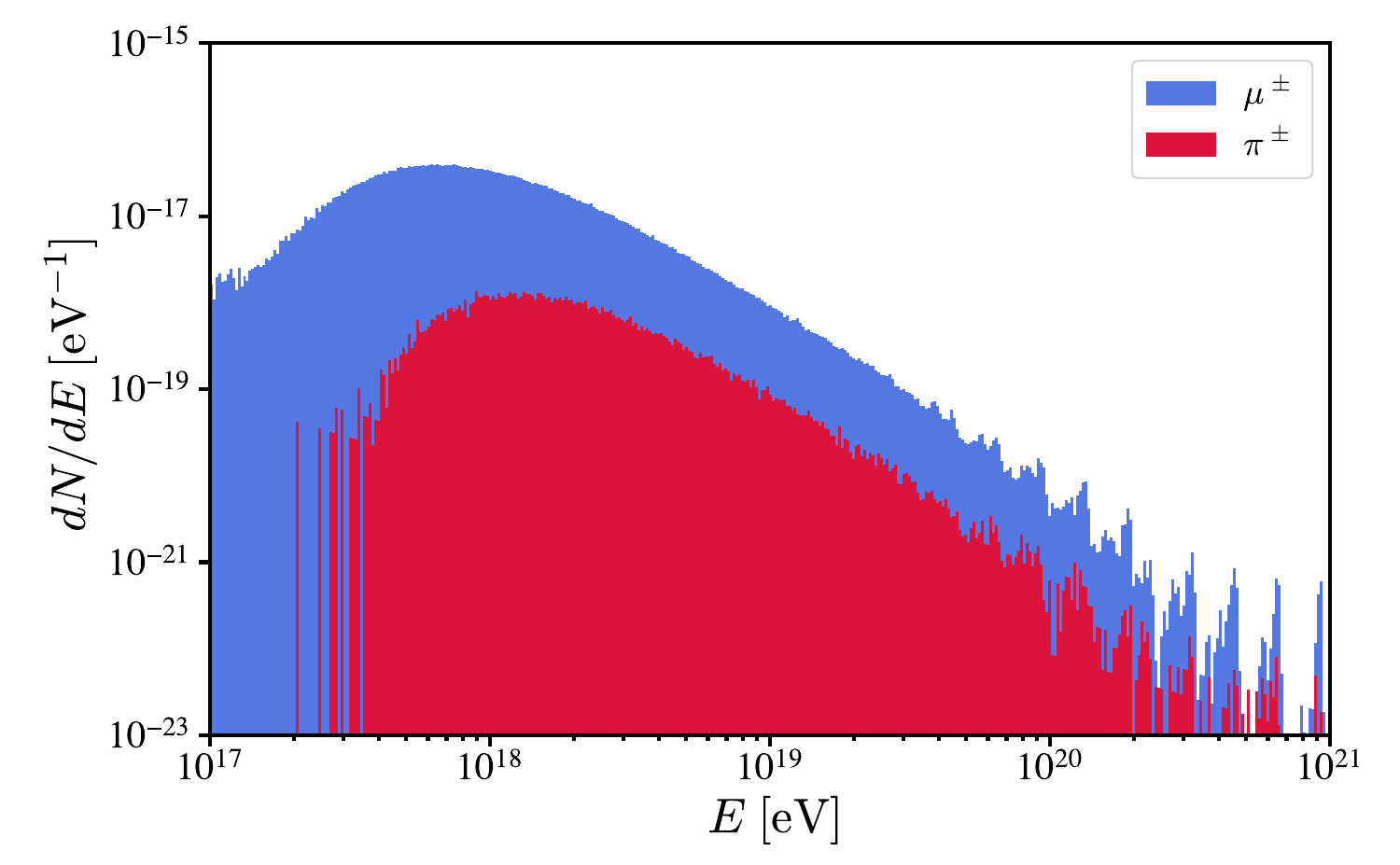}
    \caption{The spectra of muons and pions generated in the evolution of a cascade initiated by the injection of monochromatic photons with energy $E_\gamma=10^{21}$~eV at redshift $z=10$.}
    \label{fig:mu-pi-spec}
\end{figure}

The occurrences of muon and pion producing processes (that is, MPP, EMPP and CPPP) as well as their spectra can be obtained from the corresponding output files. For a photon of energy $E_\gamma=10^{21}$~eV injected at $z=10$, the occurrences and spectra of muons and pions are shown respectively in figures~\ref{fig:num_of_interactions} and \ref{fig:mu-pi-spec}. The multiple muon production in the cascade evolution can be inferred from figure~\ref{fig:num_of_interactions} with the dominant process being MPP, while the EMPP and CPPP have subleading contributions, respectively $\sim~37$\% and $\sim~8$\% with respect to MPP. The resemblance between muons and pions spectra in figure~\ref{fig:mu-pi-spec} is a consequence of treating the CPPP within the Born approximation (see~\cite{Brodsky:1971ud}) where the pion is dealt with as a structureless particle and practically as a heavier, scalar version of the muon.

\begin{figure}[t!]
    \centering
    \includegraphics[width=0.49\textwidth]{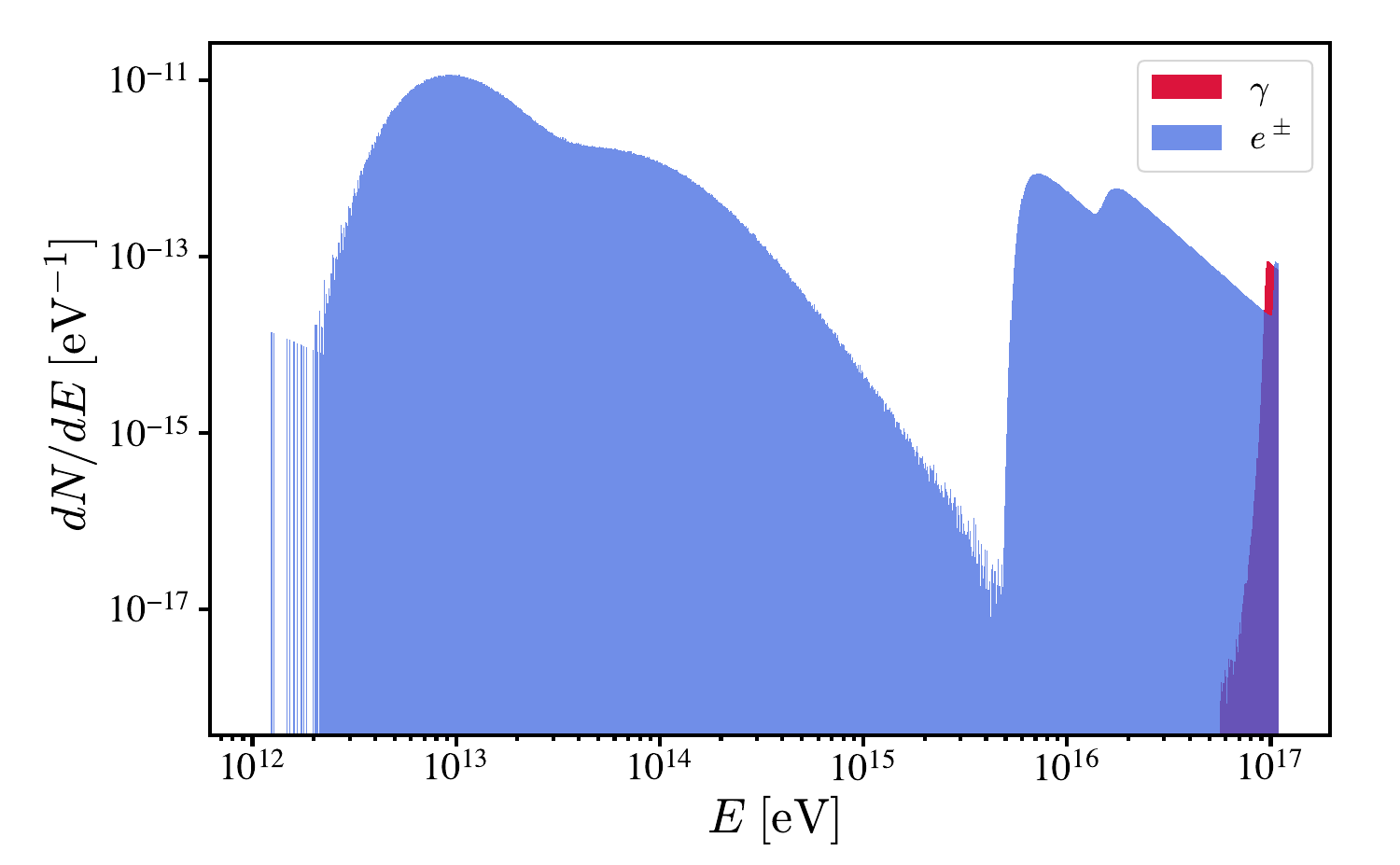}
    \caption{The electrons/positrons and photons spectra below the break energy $1.1 \times 10^{17}$~eV for monochromatic photons injected at redshift $z=10$ with energy $E_\gamma = 10^{21}$~eV (the case for the solid blue curve in figure~\ref{fig:mono_nuSpec}).}
    \label{fig:eg-spec}
\end{figure}

The spectra of electrons and photons generated below the chosen break energy $1.1 \times 10^{17}$~eV are shown respectively by the blue and red histograms in figure~\ref{fig:eg-spec}. These spectra can be obtained from \texttt{Electron\_Output} and \texttt{Gamma\_Output} files, that contain the number of electrons and photons per bin, by normalizing to the number of injected photons and dividing by the bin size. The extension of the electron spectrum to lower energies, in comparison with photon one, is mainly a consequence of DPP process and its large inelasticity while the electrons from muon decay and EMPP also contribute. Let us emphasize that the inclusion of the ETP process, while leaving almost unchanged the neutrino yield from the cascade evolution, does have an effect on the spectrum of $e^\pm$ at low energies, $E\lesssim{\rm few}\times 10^{15}$~eV, since the $e^\pm$ produced via ETP land in this range.

\subsection{Cascade inside the source: NGC 1068\label{sec:NGC1068}}

\begin{figure}[t!]
    \centering
    \includegraphics[width=0.49\textwidth]{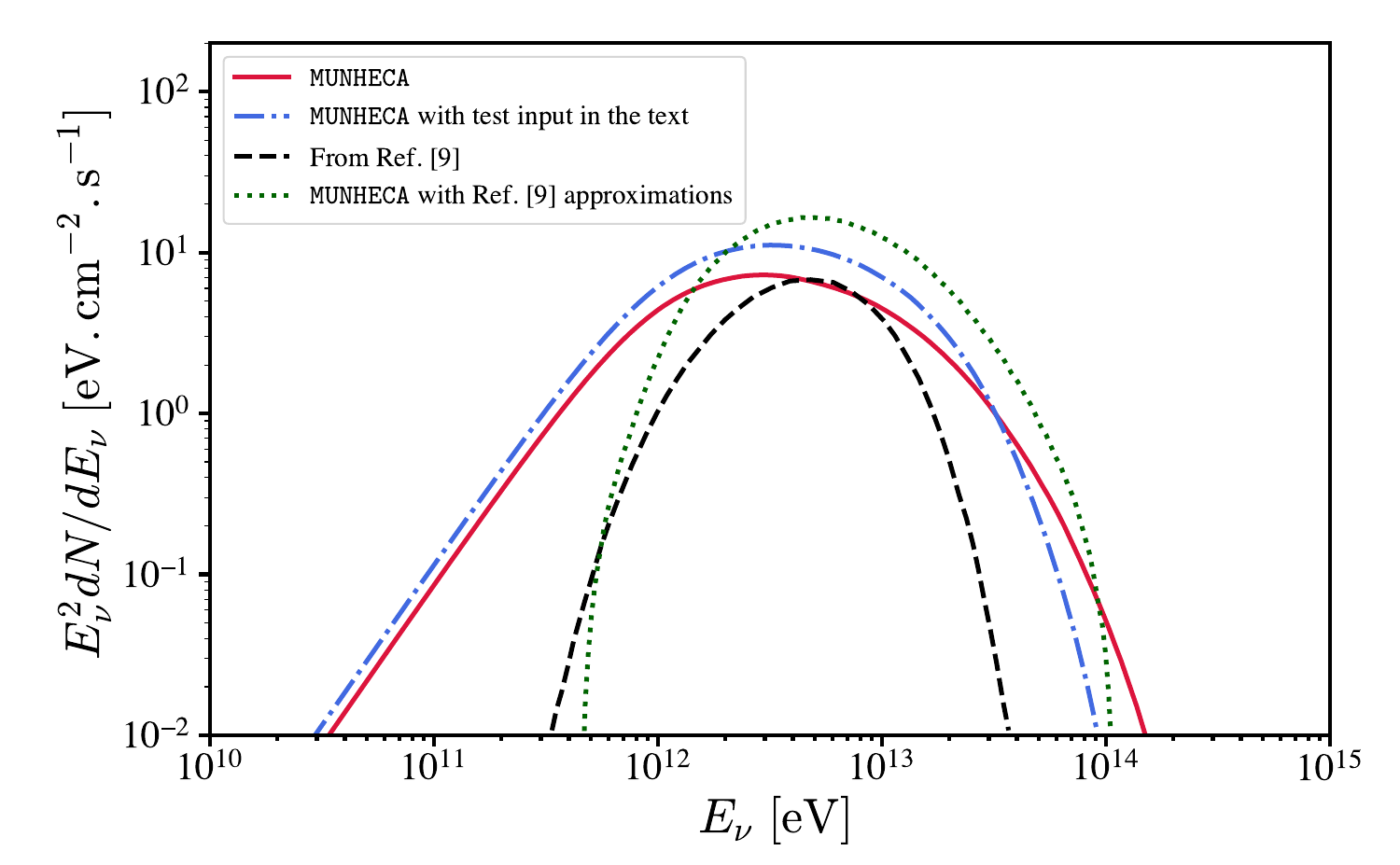}
    \caption{Neutrino yield in the interaction of a spectrum of high energy photons with target photons of black-body distribution. The result of \code{} is shown by the solid red curve and compared with the dashed black curve taken from~\cite{DanHooper:2023dls}. The dotted green curve shows an attempt to reproduce the dashed black curve by modifying \code{} to incorporate the adopted approximations of~\cite{DanHooper:2023dls} (see the text for details). The dot-dashed blue curve depicts the neutrino spectrum for the input file discussed in the text.}
    \label{fig:NGC1068}
\end{figure}

As an example of cascade development inside an astrophysical source, here we consider the recent proposal of neutrino production in the leptonic model of NGC 1068~\cite{DanHooper:2023dls}. The high energy photons, which initiate the cascade, are assumed to have a power-law spectrum with a high energy cutoff dictated by the production mechanism of these photons. For the scenario where high energy photons are produced via synchrotron radiation of high energy electrons, as is shown in~\cite{DanHooper:2023dls}, the spectrum takes the form $f(E_\gamma) = E_\gamma^{-1.5}\exp\left[\sqrt{2\pi m_e^3E_\gamma/eB}/(300\,{\rm TeV})\right]$, where $m_e$ and $e$ are respectively the electron's mass and charge, and $B=5$~kG is the magnetic field intensity at the production site. The target photons are the X-rays in the hot corona with a thermal (black-body) distribution with the temperature $T_X=1$~keV. For this temperature, the MPP can happen for $E_\gamma\gtrsim10$~TeV and hence we choose a sharp low energy cutoff on $f(E_\gamma)$ equal to 1~TeV. The form of $f(E_\gamma)$ does not match to any of the predefined spectra described in section~\ref{subsec:inout} and the method explained in section~\ref{sec:nu} should be applied. In this line, the energy range of $(1-300)$~TeV is divided into 40 bins (logarithmically spaced) and $2,000$ photons are injected in each bin. The resulting neutrino spectrum can be calculated according to eq.~(\ref{eq:weight}) and is shown by the solid red curve in figure~\ref{fig:NGC1068}. For comparison, the computed neutrino spectrum in~\cite{DanHooper:2023dls} is depicted by dashed black curve. As in~\cite{DanHooper:2023dls} the curves of figure~\ref{fig:NGC1068} are normalized such that the total power injected above 1 TeV is $1.2\times10^{43}$ erg/s. Although the overall shapes of the neutrino spectra from~\cite{DanHooper:2023dls} and \code{} are grossly in agreement near the peak of the flux, a few remarks on their differences are in order. The computation of neutrino spectrum in~\cite{DanHooper:2023dls} is based on a few approximations: the DPP, CPPP and EMTP are not taken into account, the inelasticity in MPP is taken to be equal to $50\%$, multiple muon production is neglected and the neutrino spectrum in muon decay is approximated as monochromatic. To roughly mimic this setup by \code, we turn off all the processes except MPP and modify the MPP's inelasticity and the produced neutrino spectrum in muon decay. In this way, all the approximations of~\cite{DanHooper:2023dls} are imitated except the muon production multiplicity and the result is shown by the dotted green curve in figure~\ref{fig:NGC1068} which resembles indeed more closely the dashed black curve. We conclude that, while simplified calculations might lead to correct order of magnitude results near the peak of the spectrum, especially in the low- and high-energy tails a more precise treatment, such as allowed by \code{}, is essential for reliable predictions. 

To provide an example of the input file for cascade evolution inside an astrophysical source, let us consider the power-law spectrum with exponential cutoff $E_\gamma^{-1.5}\exp[-E_\gamma/(20\,{\rm TeV})]$ which closely approximates the $f(E_\gamma)$ of~\cite{DanHooper:2023dls}. The input file of \code{} for this spectrum is:  

\texttt{Source: BlackBody}

\texttt{Temperature: 1e+3}

\texttt{BB\_E\_min: 1.0}

\texttt{BB\_E\_max: 4.e+4}

\texttt{Redshift: 0.003}

\texttt{INJ\_Spectrum: PowerLaw}

\texttt{BREAK\_Energy: 1.e+11 } 

\texttt{INJ\_Emin: 1e+12}

\texttt{INJ\_Emax: 3e+14}

\texttt{INJ\_SPEC\_Index: 1.5}

\texttt{EXP\_HIGH\_CUTOFF: ON}

\texttt{EXP\_LOW\_CUTOFF: OFF}

\texttt{ExpCut\_HighEnergy: 2.e+13}

\texttt{PHOTON\_PER\_BIN: 2000}

\texttt{NUMBER\_OF\_BINS: 40}
\\
where the range $(1\,{\rm eV}-40\,{\rm keV})$ for black-body distribution of target photons is considered and the break energy is set to $0.1$~TeV. For this input the runtime on Intel i7 processor is $\sim 26$ hours. The output of \code{} from the above input is shown by the dot-dashed blue curve in figure~\ref{fig:NGC1068}.  

Before closing this section, let us briefly comment on the scenario proposed in~\cite{DanHooper:2023dls}. If intended as a ``one zone model'', in the case of high energy photon production via synchrotron radiation, the required large magnetic field $\sim$~kG will dramatically change the cascade evolution and hence the neutrino yield. All the $e^\pm$ during the cascade evolution suffer energy loss through synchrotron radiation before making ICS or EMPP, which effectively shortens the energy loss length of leading particles and consequently suppress the muon and neutrino production. Thus, for a consistent treatment, the effect of magnetic field on cascade development should be taken into account; we are considering elaborating on it in a future update of \code. For our current purpose, the neutrino spectrum depicted in figure~\ref{fig:NGC1068} is valid for scenarios where the spectrum $f(E_\gamma)$ of high energy photons is generated by a mechanism different than synchrotron radiation, such as the ICS of high energy electrons on low energy target photons~\cite{DanHooper:2023dls}; or, it can be thought of as resulting from an effective multi-zone model, where the magnetic field threading the target photon environment is negligibly low compared to the photon source one.

\section{Discussion and Summary\label{sec:conc}}

In the course of the evolution of ultra high energy electromagnetic cascades, a fraction of energy drains into the neutrino channels, providing a new opportunity in probing the high-redshift and high energy universe via a multimessenger approach. The cascade can occur either inside an astrophysical source, where the high energy photons/electrons interact with the radiation in the source, or along the cosmological propagation of ultra high energy photons/electrons interacting with the CMB photons. The former offers a purely leptonic model for neutrino production inside the sources, while the latter is an alternative source of the ultra high energy diffuse neutrinos beside the conventional hadronic processes leading to cosmogenic neutrinos. In both cases, the generated neutrino flux, either from a population of sources or a single source, can be within the reach of present or near future neutrino observatories such as IceCube-Gen2~\cite{IceCube-Gen2:2021rkf} and GRAND~\cite{GRAND:2018iaj}. Thus, it is timely to assess the cascade evolution more thoroughly by including the important micro-physics of muon and pion producing processes. 

We introduced \code{}, a public code in \texttt{python3}, which facilitates the computation of neutrino yield in the cascade evolution, accounting for the EPP, ICS, MPP, CPPP, DPP and EMPP processes. In addition, the last two processes have been re-evaluated more accurately, as described in the appendices. The code structure and options, which include the control over the injection spectra, the choice of background photon field and the processes accounted for, are described in detail in the text. These options make \code{} a cascade simulator that can be applied to a variety of standard or exotic scenarios. 
As a concrete example of an astrophysical scenario worth exploring, we can mention the (still largely uncertain) birth and assembly processes of Super Massive Black Holes (SMBHs) at high redshift~\cite{Inayoshi:2019fun,Woods:2018lty}. These objects are notoriously difficult to study, but a peculiar ultra-high energy neutrino flux may offer another handle to constrain or detect them. 
As an example of exotic processes, we can think of the neutrino and gamma-ray fluxes associated to the decay of Super Heavy Dark Matter (SHDM) particles, notably its extragalactic flux component. In these scenarios the unstable dark matter particles can have masses exceeding $m_\chi\gtrsim 10^{18}$~eV and the processes considered here may be relevant, in particular when the dark matter is mostly or exclusively coupled to leptons, but have not been considered in the literature so far.

Compared with earlier and more simplified computations of the neutrino yield, the neutrino flux from \code{} is markedly different especially in the low-energy tail of the spectrum, which is however of utmost importance since typically easier to explore experimentally.  We believe that the current version of the code is enriched enough for it to be realistic at least in applications to high-$z$  diffuse fluxes in the baseline cosmological scenarios, as discussed in ref.~\cite{Esmaeili:2022cpz}.
Further improvements in the \code{} may be envisaged. One direction is to include a better treatment of the pion production mechanisms, going beyond Born approximation and including for instance hadronic resonances. Another avenue is to extend the cosmological cascade treatment to cases where an exotic magnetic field of cosmological origin is present: We anticipate diminishing the importance of muon-rich cascades in this case, but it would be interesting to establish the parameter space. Further, one may think of treating the cascade development in 3D: This is not particularly important for the currently considered unmagnetized cosmological application or single zone astrophysical modeling, but would be crucial if including a magnetic field, notably for multi-zone astrophysical scenarios, where the charged particle acceleration region and the secondary production one are spatially distinct and the high-energy particle flux is anisotropic. These extensions would further benefit of cross-validation with other existing codes accounting for cascade processes in the presence of a magnetic field, such as \texttt{PRESHOWER}~\cite{Homola:2013sya}. 

\begin{acknowledgments}

A.F.~E. thanks Alexander Pukhov for the help on using \texttt{calcHEP}. A.F.~E. acknowledges support by the Fundação Carlos Chagas Filho de Amparo à Pesquisa do Estado do Rio de Janeiro (FAPERJ) scholarship No. 201.293/2023, the Conselho Nacional de Desenvolvimento Científico e Tecnológico (CNPq) scholarship No. 140315/2022-5 and by the Coordenação de Aperfeiçoamento de Pessoal de Nível Superior (CAPES)/Programa de Excelência Acadêmica (PROEX) scholarship No. 88887.617120/2021-00. A.~E. thanks partial financial support by the Brazilian funding agency CNPq (grant 407149/2021). P.~D.~S. acknowledges support by the Universit\'e Savoie Mont Blanc grant {\it NoBaRaCo}. This study was partially financed by the CAPES-PRINT program No. 41/2017.
\end{acknowledgments}

\appendix

\section{Inelasticity\label{append:inelasticity}}

A key concept utilized several times in this article is the quantity $\eta$ known as `\textit{inelasticity}'. One can define $\eta^{(i)}$ as the fraction of the incoming leading particle energy ($E_{\rm in}^{(L)}$)  carried by outgoing particle $i$ ($E_{\rm out}^{(i)}$). The ``leading particle'' $L$ is the one having maximal energy in the Lab frame, i.e. $E_{\rm in}^{(L)}={\rm max}_j\{E_{\rm in}^{(j)}\}$; the concept is obviously meaningful only in the limit where $E_{\rm in}^{(L)}\gg E_{\rm in}^{(i)}$, hence the inelasticity is meaningful in the Lab frame. We define $\eta^{(i)}$ as 
\begin{equation}\label{eq:inelast_lab}
    \eta^{(i)}(s)=\frac{1}{\sigma}\int^{E_{\rm in}^{(L)}(s)}_{0}\frac{E_{\rm out}^{(i)}}{E_{\rm in}^{(L)}}\,\frac{\d\sigma}{\d E_{\rm out}^{(i)}}\,\d E_{\rm out}^{(i)}\,,
\end{equation}
where $s$ is  the squared Center of Momentum (CoM) energy, and $\d\sigma/\d E_{\rm out}^{(i)}$ is the cross section of the process, differential with respect to the $i$-th particle energy. In case where the target particles in the Lab frame are at rest, $s$ and $E_{\rm in}^{(L)}$ can be used equivalently to completely fix the kinematics. If the target particles have a non-trivial momentum distribution, an implicit average over their momentum directions is meant. The total inelasticity is then defined as $\eta\equiv \sum_{i\neq L}\eta^{(i)}$, while in the limit $E_{\rm in}^{(L)}\gg E_{\rm in}^{(i)}$ one obviously has $\sum_{i}\eta^{(i)}=1$. 

Often, the differential cross section is available in the CoM frame, while $E_{\rm in}$, $E_{\rm out}$ are known in the Lab frame; in this case, a Lorentz transformation should be performed. For the processes in which the final particles are scattered mostly in the forward/backward directions, like the ones of our interest in this article, one can write
\begin{equation}
    \frac{\d\sigma}{\d E_{\rm out}}\,\d E_{\rm out} = \frac{\d\sigma}{\d E^\ast_{\rm out}}\,\d E^\ast_{\rm out}~,
\end{equation}
where the CoM quantities are marked by `$\ast$'. When the scattered particles are strongly collimated in forward/backward directions, the Lorentz transformation can be approximated by $E_{\rm out} = E_{\rm out}^\ast\gamma_c(1\pm\beta_{\rm out}^\ast)$, where $\beta_{\rm out}$ is the velocity of the outgoing particle, $\gamma_c = E_{\rm out}/\sqrt{s}$ is the boost factor from the CoM to lab frame, and the $+(-)$ sign designates the forward (backward) direction. Thus, in the high energy regime ($\beta_{\rm out}\to1$) the inelasticity of the forward scattered particles, $\eta_+$, can be obtained by using $E_{\rm out} = 2\gamma_cE_{\rm out}^\ast$,   
\begin{equation}\label{eq:eta_forward}
    \eta_{+}(s) = \frac{1}{\sigma}\int_{0}^{\sqrt{s}/2}\frac{2E_{\rm out}^\ast}{\sqrt{s}}\frac{\d\sigma}{\d E_{\rm out}^\ast}\,\d E_{\rm out}^\ast~.
\end{equation}
For the backward particles, 
\begin{equation}
E_{\rm out} = E_{\rm out}^\ast\gamma_c(1-\beta_{\rm out}^\ast)=\frac{E_{\rm out}^\ast\gamma_c}{(\gamma_{\rm out}^\ast)^2(1+\beta_{\rm out}^\ast)}\simeq \frac{\gamma_c m^2}{2E_{\rm out}^\ast}\,,\nonumber
\end{equation} where $m$ is the outgoing particle mass. The inelasticity for backward scattered particle, $\eta_-$, can be written as
\begin{equation}\label{eq:eta_back}
    \eta_{-}(s) = \frac{1}{\sigma}\int^{\sqrt{s}/2}_{\frac{m^2}{2\sqrt{s}}}\frac{m^2}{2\sqrt{s}}\frac{1}{E_{\rm out}^\ast}\frac{\d\sigma}{\d E_{\rm out}^\ast}\,\d E_{\rm out}^\ast~.\
\end{equation}

Using the convention of Ref.~\cite{Demidov:2008az}, the differential cross section for the outgoing particle $\ell$ can be parameterized as
\begin{equation}\label{eq:phi}
    \frac{{\rm d}\sigma_\ell}{{\rm d}E^\ast}(s) = \frac{1}{\sqrt{s}}\,\phi_\ell(r_\ell,s)\sigma_{\rm tot}(s)~,
\end{equation}
where $r_\ell=2E_{\ell,{\rm out}}^\ast/\sqrt{s}$ and $\phi_\ell(r_\ell,s)$ is a function that should satisfy 
$$\int_{0}^{1}\phi_\ell(r_\ell,s)\,\d r_\ell = 2 \quad {\rm and} \quad \sum_\ell\int_0^1 r_\ell\phi_\ell(r_\ell,s)\,\d r_\ell =4~,$$
respectively imposing the conservation of probability and energy. Using the function $\phi_\ell(r_\ell,s)$ one can calculate the inelasticity of particle $\ell$, scattered either in forward or backward direction, by  rewriting respectively the eqs.~(\ref{eq:eta_forward}) and (\ref{eq:eta_back}) as
\begin{equation}\label{eq:eta_phi_forward}
    \eta_{\ell,+}(s) = \int_{0}^{1}\frac{r_\ell}{2}\,\phi_\ell(r_\ell,s)\,\d r_\ell~,
\end{equation}
and 
\begin{equation}\label{eq:eta_phi_back}
    \eta_{\ell,-}(s)= \int^{1}_{\frac{m^2}{s}}\frac{m^2}{2sr_\ell}\,\phi_\ell(r_\ell,s)\,\d r_\ell~.
\end{equation}

\section{Double Pair Production (DPP)\label{append:dpp}}

\begin{figure}[t!]
    \centering
    \subfloat[]{
    \includegraphics[width=0.158\textwidth]{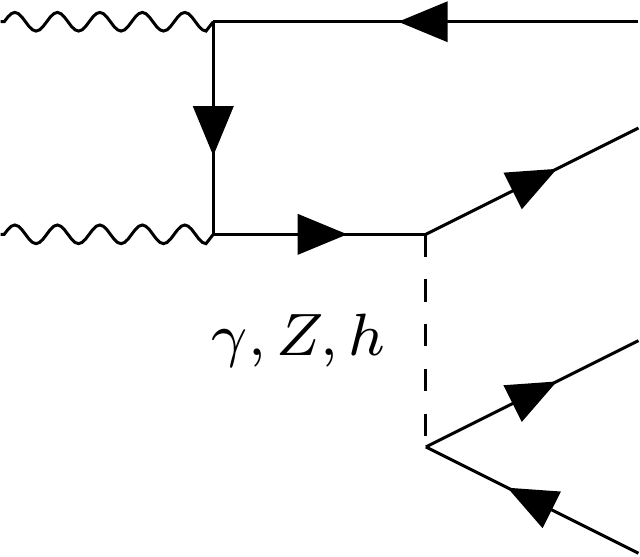}
    \label{fig:diag_a}
    }
    \hspace*{0.02\textwidth}
    \subfloat[]{
    \includegraphics[width=0.127\textwidth]{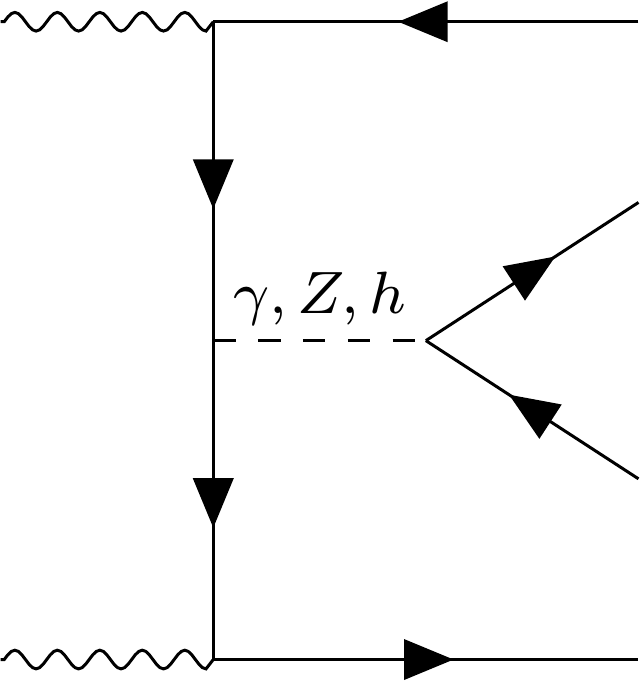}
    \label{fig:diag_b}
    }
    \hspace*{0.02\textwidth}
    \subfloat[]{
    \includegraphics[width=0.085\textwidth]{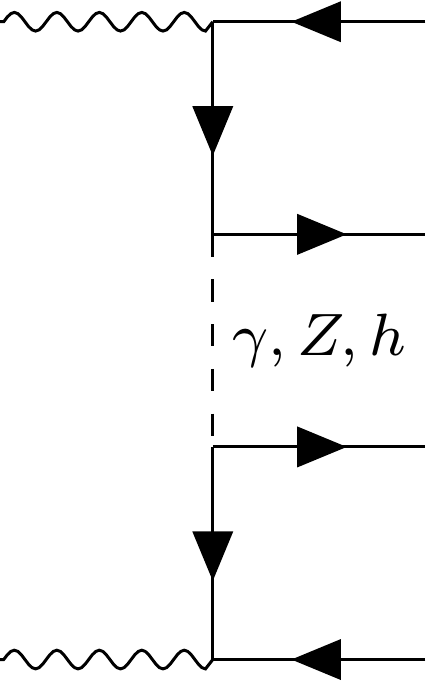}
    \label{fig:diag_c}
    }
    \caption{DPP Feynman diagrams at tree-level: Wavy lines are photons, solid lines are electrons and dashed lines are mediator propagators (see discussion in the text).}
    \label{fig:FeynmanDiags}
\end{figure}

To calculate the energy fraction carried by each of the final particles in DPP we need the differential cross sections and inelasticity of this process, which can be derived at leading order in perturbation theory from the tree level Feynman diagrams shown schematically in figure~\ref{fig:FeynmanDiags}. We use the \verb|calcHEP.3.8.10|~\footnote{\texttt{https://theory.sinp.msu.ru/\~{}pukhov/calchep.html}} code~\cite{Belyaev:2012qa} for the numerical computation of Feynman diagrams. \verb|calcHEP| provides an automatic evaluation of the matrix elements and their squares, and performs Monte Carlo phase space integration, via the \texttt{VEGAS} algorithm, for elementary particle collisions and decays at the lowest order in perturbation theory~\cite{Lepage:1977sw,Lepage:2020tgj}.

The Feynman diagrams of DPP can be arranged according to either the topology of the diagram, depicted as classes (a), (b) and (c) in figure~\ref{fig:FeynmanDiags}, or the mediator, depicted by the dashed lines, which can be the photon, the $Z$ boson or the Higgs ($h$). Among the three topologies, at high energy the main contribution to the total cross section of DPP comes from type (c), having $\frac{\sigma_a+\sigma_b}{\sigma_c}|_{s=100{~\rm GeV^2}}\approx 10^{-5}$, while at energies close to the threshold of DPP the types (a) and (b) diagrams become relevant such that $\frac{\sigma_a+\sigma_b}{\sigma_c}|_{s=10^{-4}{~\rm GeV^2}}\approx 0.17$. Here $s$ is the center of momentum energy squared. Among the three mediators, the contributions of $Z$ and $h$ mediators are negligible, respectively $\mathcal{O}(10^{-12})$ and $\mathcal{O}(10^{-32})$ with respect to photon-mediator diagrams. Thus, for the rest of our discussion, we consider only the diagrams with the photon mediator.

\begin{figure}[t!]
    \centering
    \includegraphics[width=0.48\textwidth]{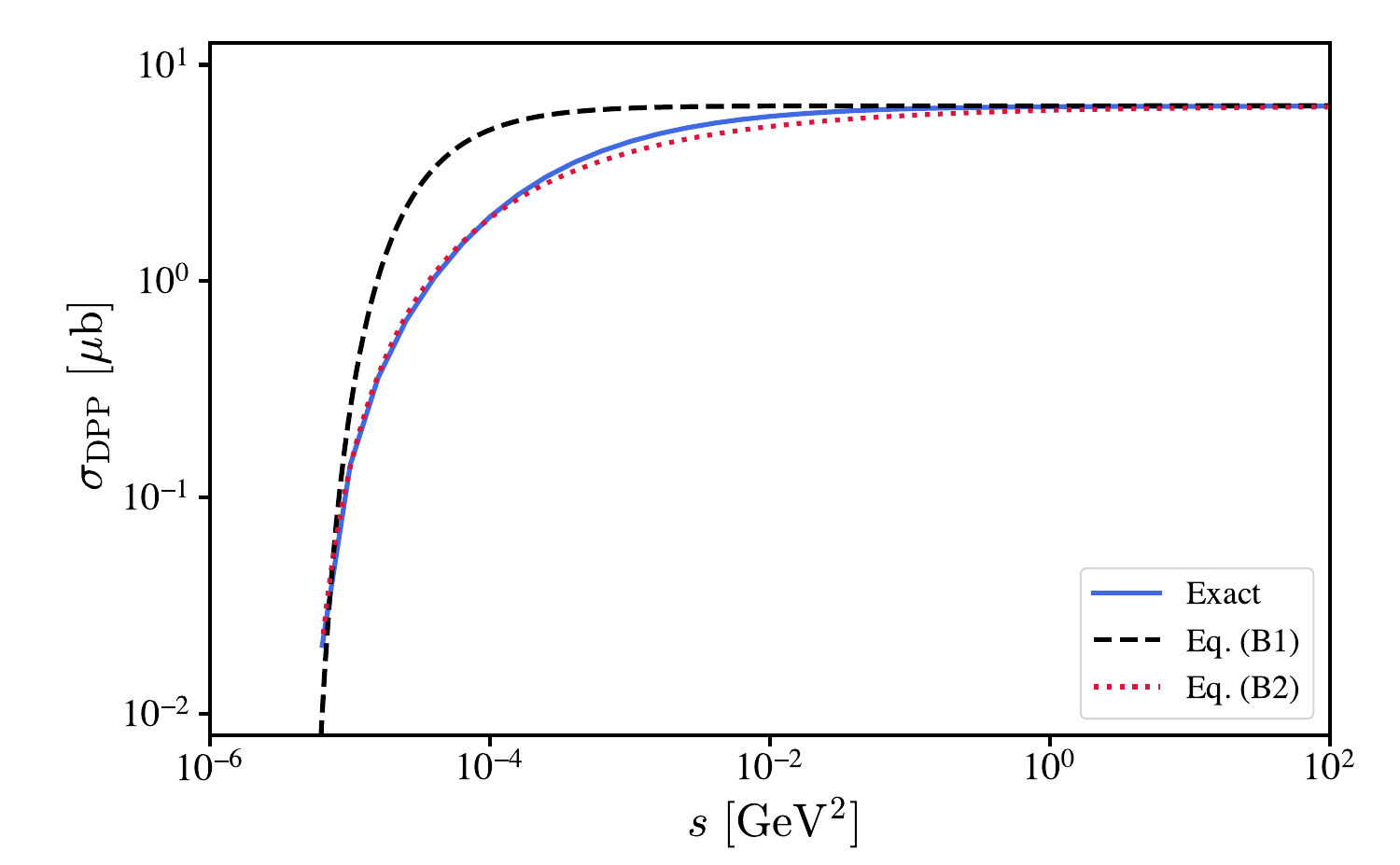}
    \caption{Total cross section of DPP as function of $s$: the result of numerical computation is shown by the blue solid curve, the black dashed and red dotted curves show respectively the approximations in eq.~(\ref{eq:dpp_approx}) and eq.~(\ref{eq:fittotxs}).}
    \label{fig:DPP_tot}
\end{figure}

The following approximation of the total cross section of DPP, based on a fit to the numerical integration over the phase space close to the threshold, is reported in the literature~\cite{Cheng:1970_DPP,galanti:2020ea,Brown:1973onk}
\begin{equation}\label{eq:dpp_approx}
    \sigma_{\rm DPP}(s)\simeq \frac{\alpha^4\theta(s-s_{\rm th})}{\pi m_e^2}\left(\frac{175}{36}\zeta(3)-\frac{19}{18}\right)\left(1-\frac{s_{\rm th}}{s}\right)^6~,
\end{equation}
where $\alpha$ is the fine structure constant, $m_e$ is the electron mass, $\zeta(3)=1.202$ is the Riemann zeta function and the threshold value $s_{\rm th}=16m_e^2$ is taken into account by the step function $\theta$. At high energies, eq.~\eqref{eq:dpp_approx} approaches the asymptotic value $\sigma_{\rm DPP}(s\to\infty)\approx 6.45\,\mu$b. To assess the validity of eq.~\eqref{eq:dpp_approx}, in figure~\ref{fig:DPP_tot} we compare it (dashed black curve) with the numerical evaluation of $\sigma_{\rm DPP}$ (solid blue curve): Deviations are evident below $s\lesssim 10^{-2}$~GeV$^2$. A better approximation can be obtained by the following expression 
\begin{equation}\label{eq:fittotxs}
    \sigma_{\rm DPP}(s)\simeq 6.45\,\mu{\rm b}\left[1-\left(\frac{s_{\rm th}}{s}\right)^{1/3}\right]^{14/5}~,
\end{equation} 
which is depicted by the dotted red curve in figure~\ref{fig:DPP_tot}. Eq.~(\ref{eq:fittotxs}) approximates the exact DPP cross section better than $\sim 10\%$ at all values of $s$, while eq.~(\ref{eq:dpp_approx}) deviates from the exact cross section by a factor $\sim$ two.

\begin{figure}[t!]
    \centering
    \includegraphics[width=0.49\textwidth]{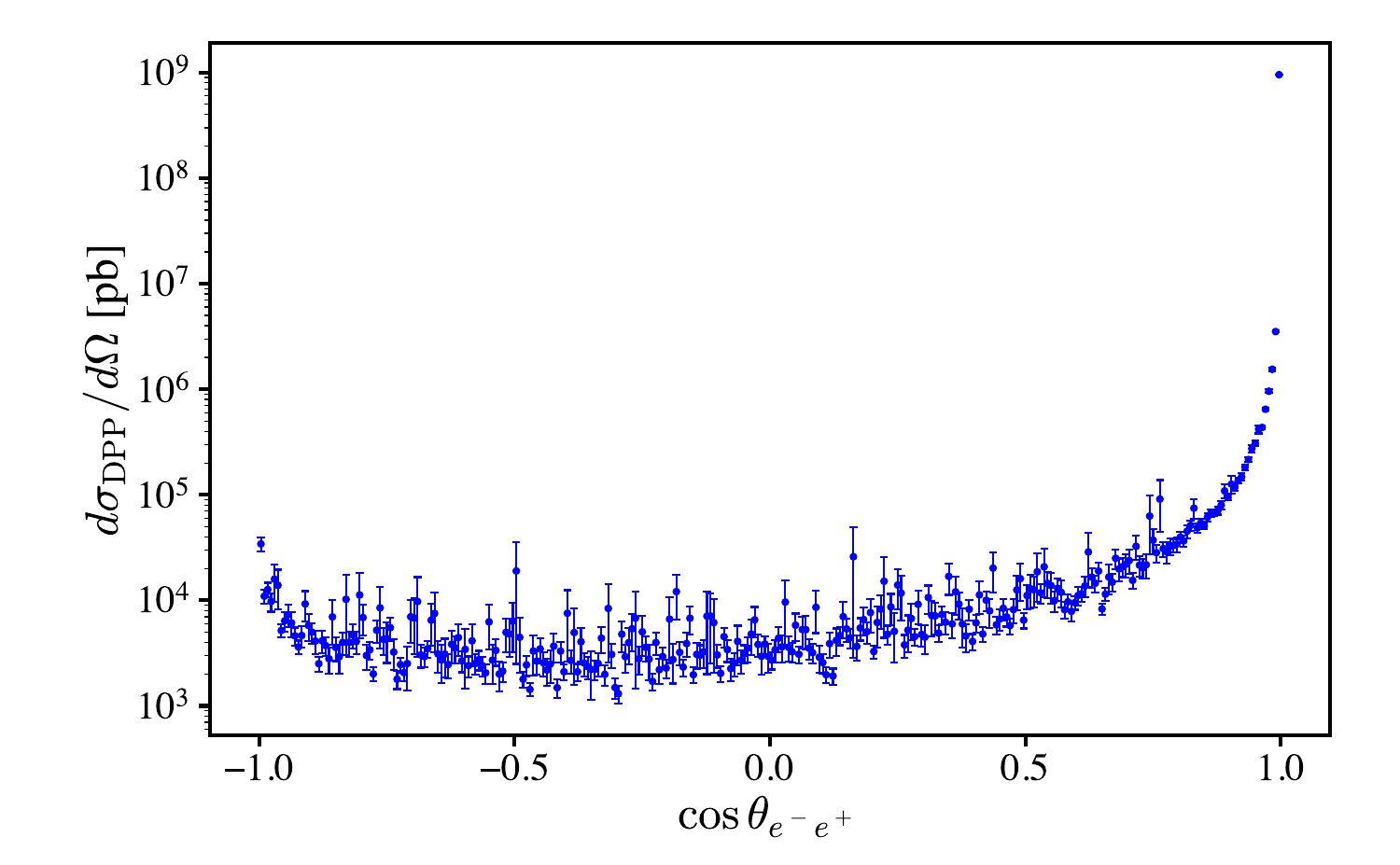}
    \caption{The differential cross section $\d\sigma_{\rm DPP}/\d\Omega$ in the CoM frame, as function of the angle between the electron and positron in a pair, $\cos\theta_{e^-e^+}$, for $s=100~{\rm GeV^2}$.}
    \label{fig:angles1}
\end{figure}

The angular distribution of the outgoing electrons and positrons in DPP can be inferred from the differential cross section. 
In the CoM frame, the four outgoing electrons/positrons are typically emitted `back to back' in pairs,  i.e. each electron is closest in angular space to a positron, and the angular separation between $e^+$ and $e^-$ is small; the two pairs $e^\pm$ are  separated by an angle $\pi$.  To assess the degree of accuracy of this statement, the differential cross section $\d\sigma_{\rm DPP}/\d\Omega$ in the CoM frame as a function of $\cos\theta_{e^-e^+}$ is shown in figure~\ref{fig:angles1}, where $\theta_{e^-e^+}$ in the angle between the electron and the positron. The sharp peak at $\theta_{e^-e^+}\approx0$ shows that the members in each pair are collinear with error $\sim 10^{-5}$ (at the chosen value $s=100~{\rm GeV^2}$). Close to the threshold of DPP, this statement has to be mitigated; still, only $\sim10^{-2}$ of the events escape the above-mentioned simplifying classification. 
 
Although the tree level diagrams in figure~\ref{fig:FeynmanDiags} are straightforward to compute, a technical remark is in order: To distinguish between the two $e^\pm$ pairs in DPP and at the same time to ease the convergence of phase space integrals, it is convenient to differentiate between the two pairs by assigning different names to them  (i.e. to treat them as if they were distinguishable, like if they belonged to a different lepton family) while keeping the masses of the new leptons equal to the electron mass. This, of course, would artificially double the yields, an effect which can be compensated for by setting a cut requiring positive rapidity for one of the pairs. With reference to the oriented direction of the high energy photon in the Lab frame, we label the pairs as {\it forward} and {\it backward}, respectively containing $e^\pm_{\rm F}$ and $e^\pm_{\rm B}$ electrons/positrons.  

\begin{figure}[t!]
    \centering
    \includegraphics[width=0.49\textwidth]{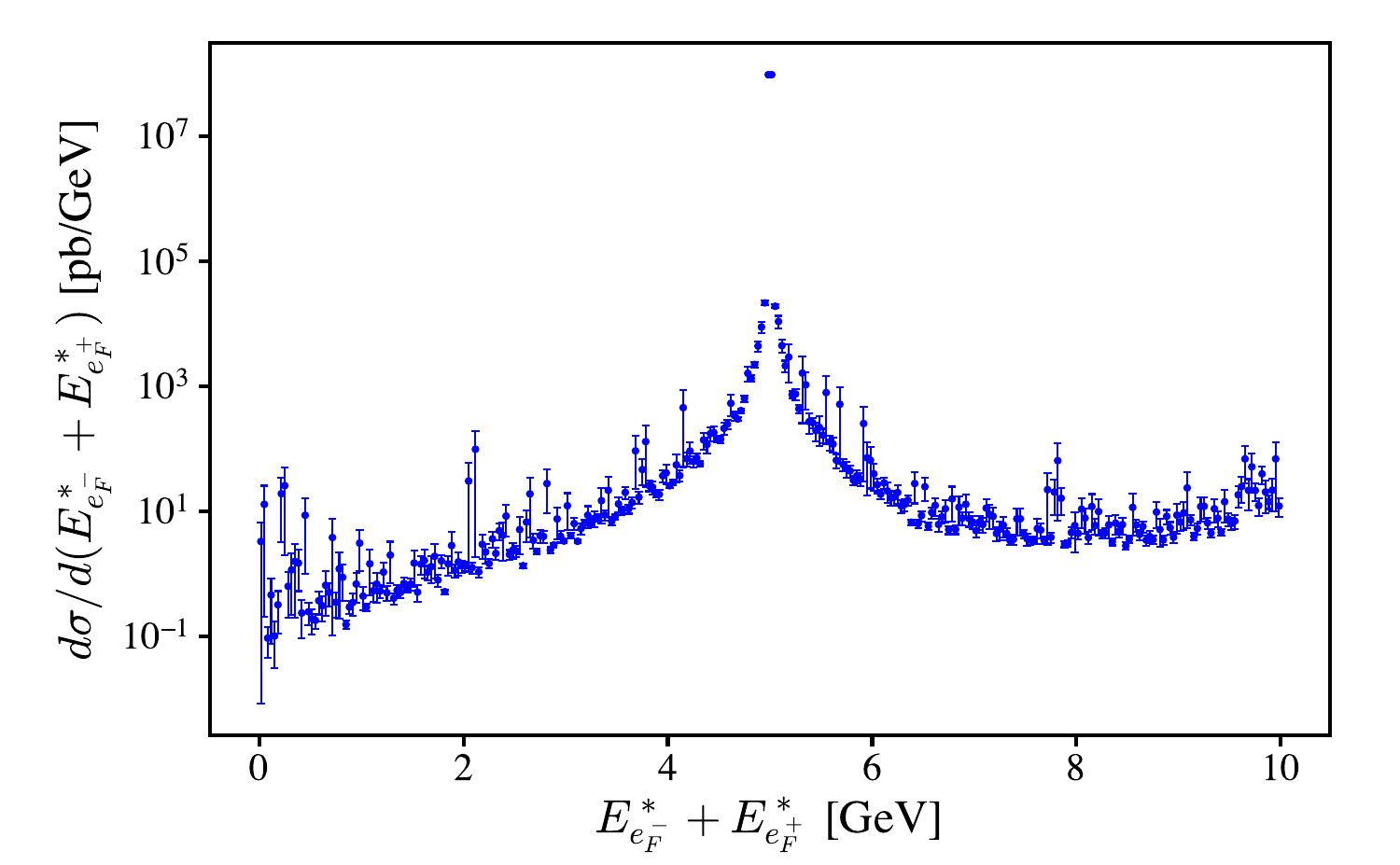}
    \caption{Differential cross section as function of the total energy of the forward pair $E^\ast_{e^-_{\rm F}}+E^\ast_{e^+_{\rm F}}$ in the CoM frame, assuming $s=100~{\rm GeV^2}$, {\it i.e.} the photons in the initial state have $p_{\gamma,1}=p_{\gamma,2}=5~{\rm GeV}$.}
    \label{fig:1D_EeE}
\end{figure}

The energy distribution between the final particles can be assessed by considering the single and double differential cross sections with respect to one or two electron energy. Here, we show the distributions for the case $s=100~{\rm GeV^2}$ in the CoM frame as a benchmark. The differential cross section with respect to the sum of $e^+_{\rm F}$ and $e^-_{\rm F}$ energies has a sharp peak at the corresponding incoming photon energy, half of the total energy (here 5~GeV), as shown in figure~\ref{fig:1D_EeE}. For the backward pair, the distribution is of course the mirror image of figure~\ref{fig:1D_EeE} with respect to the peak. In other words, the energy is shared equally between the forward and backward pairs.

\begin{figure}[t!]
    \centering
    \includegraphics[width=0.48\textwidth]{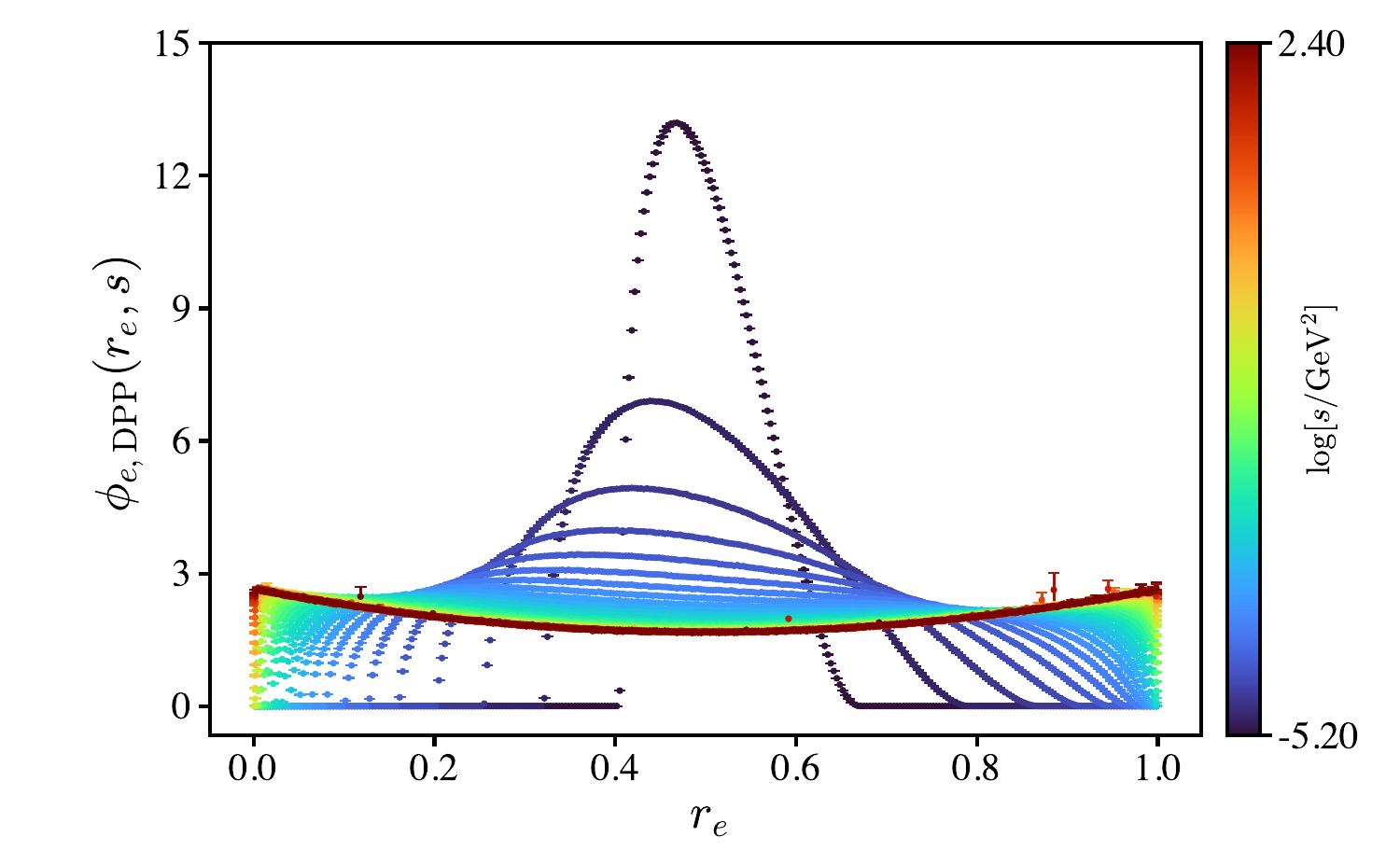}
    \caption{$\phi_{e,{\rm DPP}}(r_e,s)$ as function of $r_e=2E^\ast_e/\sqrt{s}$ and for different values of CoM squared energy $s$.}
    \label{fig:phi_dpp}
\end{figure}

\begin{figure}[t!]
    \centering
    \includegraphics[width=0.48\textwidth]{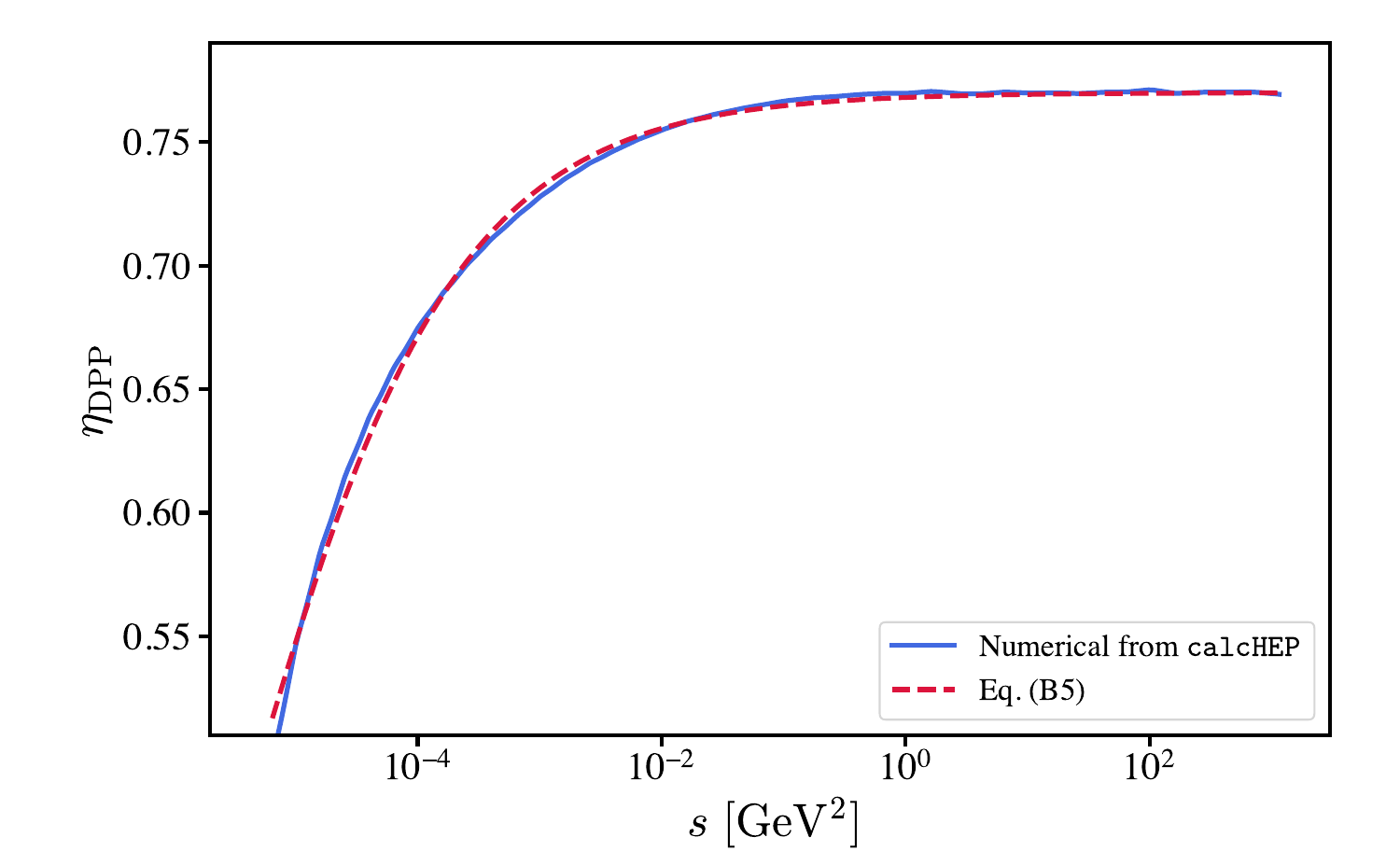}
    \caption{The inelasticity of the higher energy member of the forward pair, eq.~(\ref{eq:etaHigh_dpp}), in DPP, as function of $s$. The solid blue and red dashed curves show the result of numerical calculation and the fit of eq.~(\ref{eq:fit-dpp}), respectively. }
    \label{fig:eta}
\end{figure}

The energy partition within each pair can be understood from the differential cross section with respect to the energy of one of the members in each pair; {\it i.e.}, the function $\phi_{e,{\rm DPP}}(r_e,s)$ defined in eq.~(\ref{eq:phi}), computed by \texttt{calcHEP} and depicted in figure~\ref{fig:phi_dpp} for different values of $s$, where $r_e = 2E^\ast_e/\sqrt{s}$ is the energy fraction of one electron ($E^\ast_e$ being the electron energy) in the CoM frame. Close to the threshold of DPP, and in CoM frame, in each pair the electron and positron share the pair energy almost equally (see the peak at $r_e\approx0.5$ for the darker color scale in figure~\ref{fig:phi_dpp}). For $s\gtrsim 0.01~{\rm GeV^2}$, the energy distribution is wide and can be fitted accurately by $\phi_{\rm fit}(r)=5/3+(2r-1)^2$~\cite{Demidov:2008az}, which means that the relative energy share of the members in the pair is almost equally probable for any value from zero to one. The maximum energy shown in Figures. \ref{fig:phi_dpp} and \ref{fig:eta} ($s\sim 100{~\rm GeV^2}$) is well beyond the interesting energy range in this article. At energies even higher than those covered here one should worry about electroweak processes, but we deem that regime so extreme compared to what are currently considered realistic  scenarios that we can safely omit its treatment at this stage. Moreover, the numerical computation with \texttt{calcHEP} shows that the fit proposed for $\phi_{e,{\rm DPP}}$ can be safely  utilized at least up to $s\sim 10^4~{\rm GeV^2}$.

Using the $\phi_{e,{\rm DPP}}$ in eqs.~(\ref{eq:eta_phi_forward}) and (\ref{eq:eta_phi_back}), the {\it average} inelasticities of the backward and forward pairs can be computed. Increasing the energy, the average fraction of energy carried by the backward pair in the lab frame decreases such that almost 100\% of the leading photon energy is transferred to the forward pair. To compute the partition of energy within the forward pair, we can resort to the particles inelasticities. Because of the symmetric energy sharing in CoM frame, the inelasticities of the members of forward pair can be written as
\begin{equation}\label{eq:etaHigh_dpp}
    \eta_{H}(s) = 2\,\int_{0.5}^{1}\frac{r_e}{2}\,\phi_{e,{\rm DPP}}(r_e,s)\,\d r_e~,
\end{equation}
and
\begin{equation}\label{eq:etaLow_dpp}
    \eta_{L}(s) = 2\,\int_{0}^{0.5}\frac{r_e}{2}\,\phi_{e,{\rm DPP}}(r_e,s)\,\d r_e~,
\end{equation}
where the $H$ and $L$ respectively refer to the higher and lower energy member of the forward pair. The solid blue curve in figure~\ref{fig:eta} shows $\eta_H$ as function of $s$. A fit to this curve can be written as
\begin{equation}\label{eq:fit-dpp}
    \eta(s) \approx a+b\,\exp{[-(s_{\rm th}/s)^{c}]}~,
\end{equation}
with fit parameters $a = 0.32$, $b = 0.45$, and $c = 0.44$, and is depicted by the dashed red curve in figure~\ref{fig:eta}. From this figure, on the average the maximum energy fraction carried by the high energy member of the forward pair is $\sim 77\%$ at $s\gtrsim0.1\,{\rm GeV}^2$; {\it i.e.}, an average energy share with the ratio $1:3$ between the members of the forward pair.   

Remembering that Ref.~\cite{Demidov:2008az} claims that the forward pair takes all the initial energy and shares it equally between the members, we find that at $s\gtrsim10^{-2}~{\rm GeV}^2$ a fraction of $\sim10^{-5}$ of the leading particle energy is carried by the backward pair, thus confirming their first approximation; however, we find that---apart for near threshold---the bulk of the energy is shared between the forward pair members with an average ratio $1:3$, so the other approximation of Ref.~\cite{Demidov:2008az} is relatively poor and leads to errors of several tens of percent.

\section{Electron Muon-Pair Production (EMPP)\label{append:EMPP}}

\begin{figure}[h!]
    \centering
    \subfloat[]{
    \includegraphics[width=0.2\textwidth]{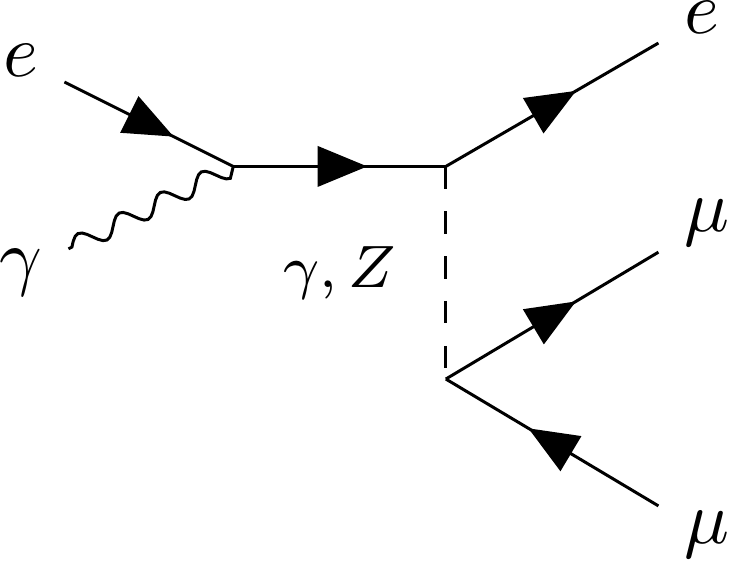}
    \label{fig:diag_EMPPa}
    }
    \hspace*{0.02\textwidth}
    \subfloat[]{
    \includegraphics[width=0.2\textwidth]{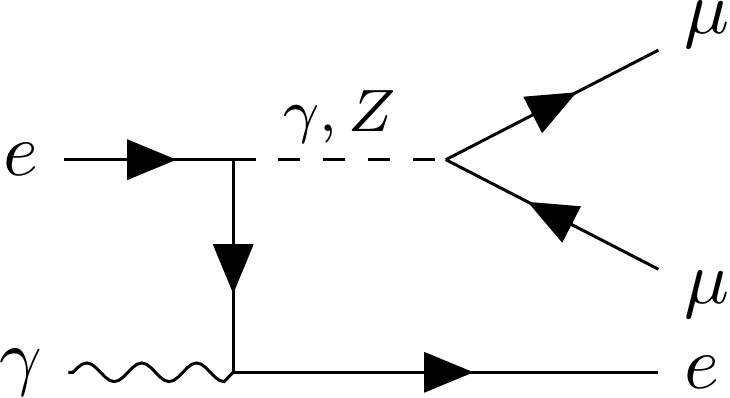}
    \label{fig:diag_EMPPb}
    }
    \newline
    \subfloat[]{
    \includegraphics[width=0.15\textwidth]{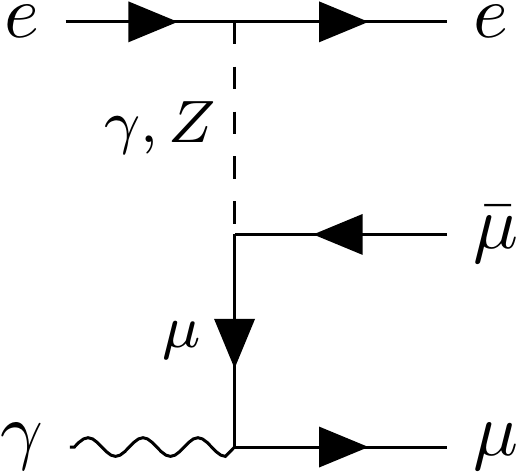}
    \label{fig:diag_EMPPc}
    }
    \hspace*{0.1\textwidth}
    \subfloat[]{
    \includegraphics[width=0.15\textwidth]{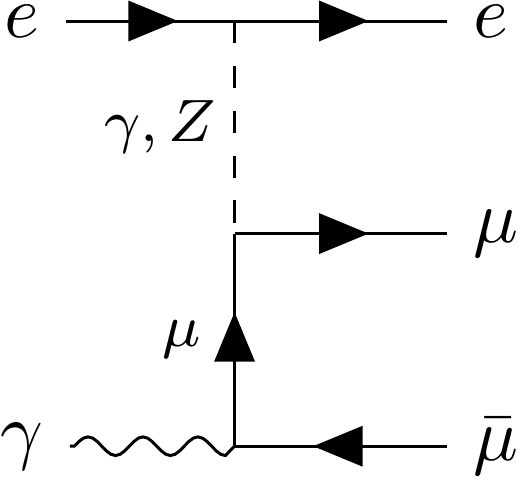}
    \label{fig:diag_EMPPd}
    }
    \caption{EMPP Feynman diagrams at tree-level: Wavy lines are photons, solid lines electrons and dashed lines are mediator propagators (see discussion in the text).}
    \label{fig:Feyn_EMPP}
\end{figure}

The four types of EMPP Feynman diagrams are depicted in figure~\ref{fig:Feyn_EMPP}. The contribution of the diagrams with $Z$ boson exchange are $\mathcal{O}(10^{-10})$ smaller than the diagrams with photon propagator. Also, the relative contributions of diagrams in figures~\ref{fig:diag_EMPPa} and \ref{fig:diag_EMPPb} to the total EMPP cross section are $\mathcal{O}(10^{-3}-10^{-2})$. The total cross section of EMPP, at high $s$ and in the range $5m_\mu^2 < s < 20m_\mu^2$ where $m_\mu$ is the muon mass, has been discussed and computed both numerically (using \texttt{compHEP}) and analytically (by equivalent-photon approximation) in~\cite{Athar:2001sm}. For this work, as a cross-check and since the total and differential cross sections are needed in a wider range of energy, we re-computed them by \texttt{calcHEP}. 

\begin{figure}[t!]
    \centering
    \includegraphics[width=0.49\textwidth]{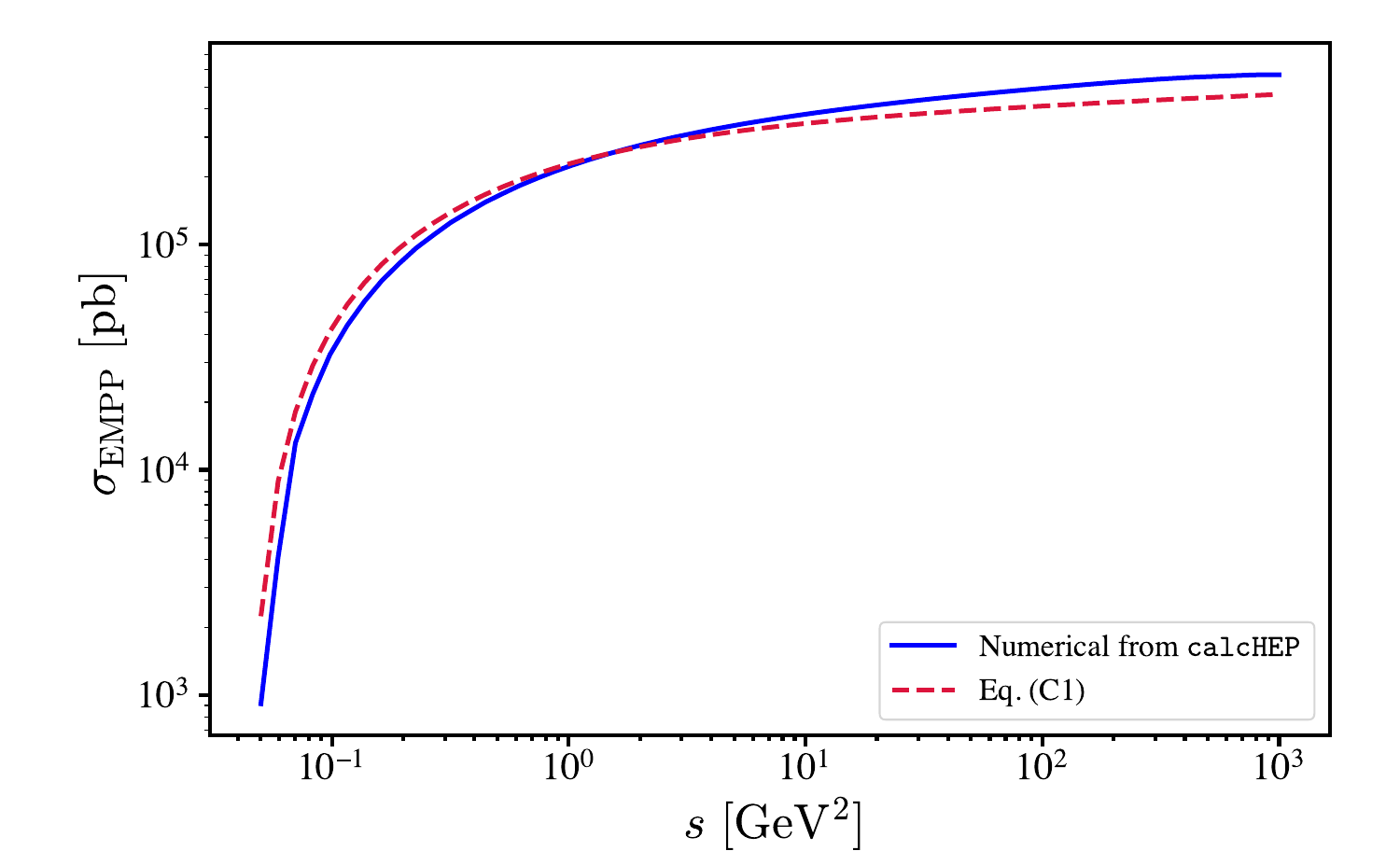}
    \caption{The total cross section of EMPP from the approximation of eq.~(\ref{eq:empp-app}) and \texttt{calcHEP}, shown respectively by the dashed red and solid blue curves.}
    \label{fig:Xsec_EMPP}
\end{figure}

In the equivalent-photon approximation, the EMPP total cross section can be estimated from the MPP cross section by~\cite{Athar:2001sm}
\begin{equation}\label{eq:empp-app}
    \sigma_{\rm EMPP}(s) \approx \int_{4m_\mu^2/s}^{1}dx\, f_{\gamma/e}(x)\,\sigma_{\rm MPP}(\hat{s}=xs)~,
\end{equation}
where $f_{\gamma/e}(x) = (\alpha/2\pi)[\left(1+(1-x)^2\right)/x]\ln(s/m_e^2)$ is the probability of the emission of a photon with energy $E_\gamma$ from the incident electron with energy $E_e$, where $x=E_\gamma/E_e$. The approximation in eq.~(\ref{eq:empp-app}) is shown by the dashed red curve in figure~\ref{fig:Xsec_EMPP} and is compared with the \texttt{calcHEP} numerical computation depicted by the solid blue curve.

\begin{figure}[t!]
    \centering
    \includegraphics[width=0.48\textwidth]{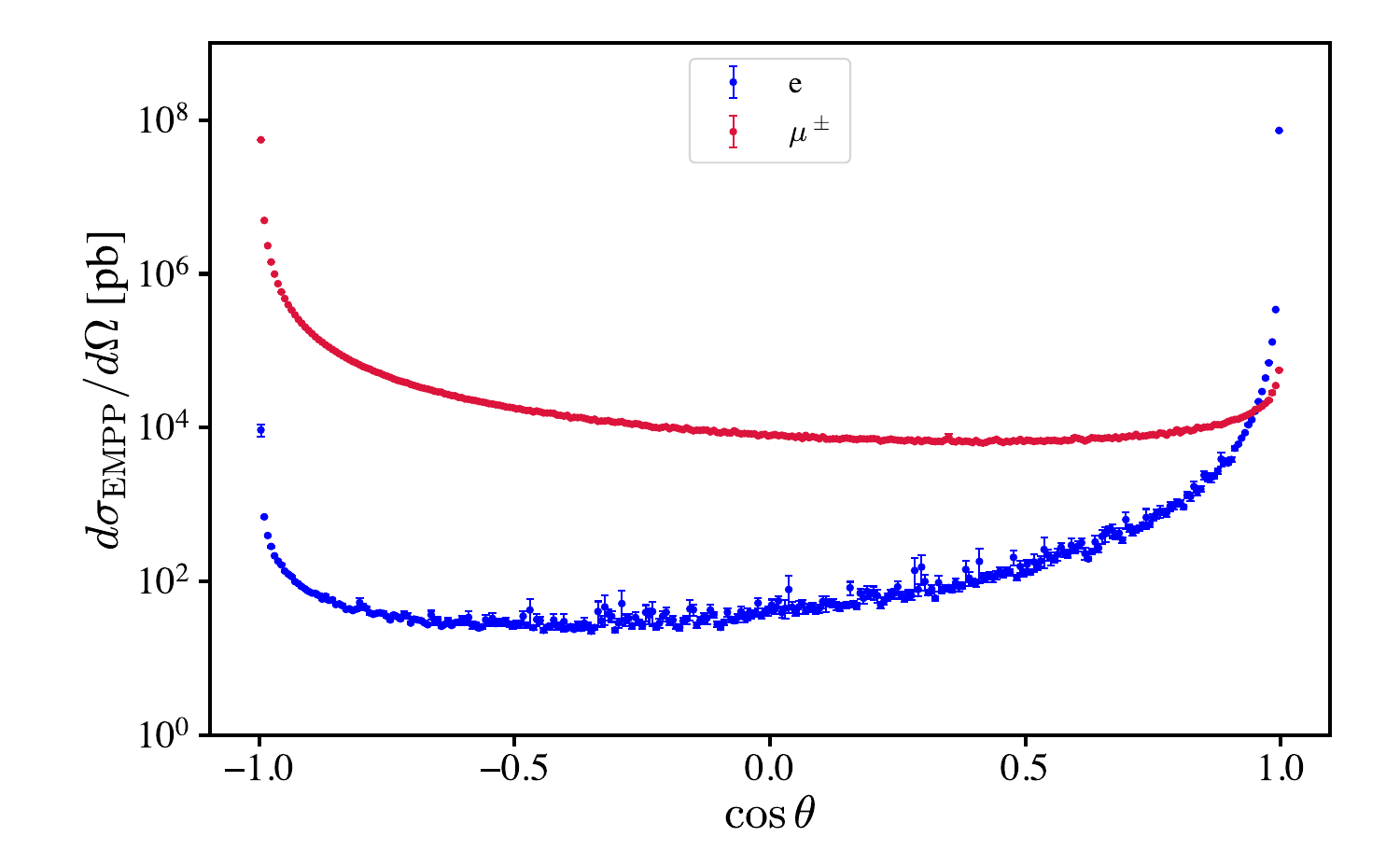}
    \caption{The angular distributions of the outgoing electron (blue color) and muons (red color) in EMPP process, at $s = 100~{\rm GeV^2}$ and in the CoM frame. $\theta$ is the angle between the direction of outgoing particle and the collision axis in the CoM frame.}
    \label{fig:EMPP_anlge1}
\end{figure}

The angular distributions of the outgoing electron and muons, at $s = 100~{\rm GeV^2}$ and in the CoM frame, are depicted in figure~\ref{fig:EMPP_anlge1} respectively by the red and blue colors, where $\theta$ is the angle between the direction of outgoing particle and the collision axis. As can be seen from this plot, with errors $<10^{-4}$ and $<10^{-3}$ the electron and muons are scattered in the forward and backward directions, respectively. Thus, once more, the forward/backward configuration of the outgoing particles helps us to calculate the inelasticity just from the energy distribution in the interaction, as explained in Appendix~\ref{append:inelasticity}.    

\begin{figure}[t!]
    \centering
    \includegraphics[width=0.5\textwidth]{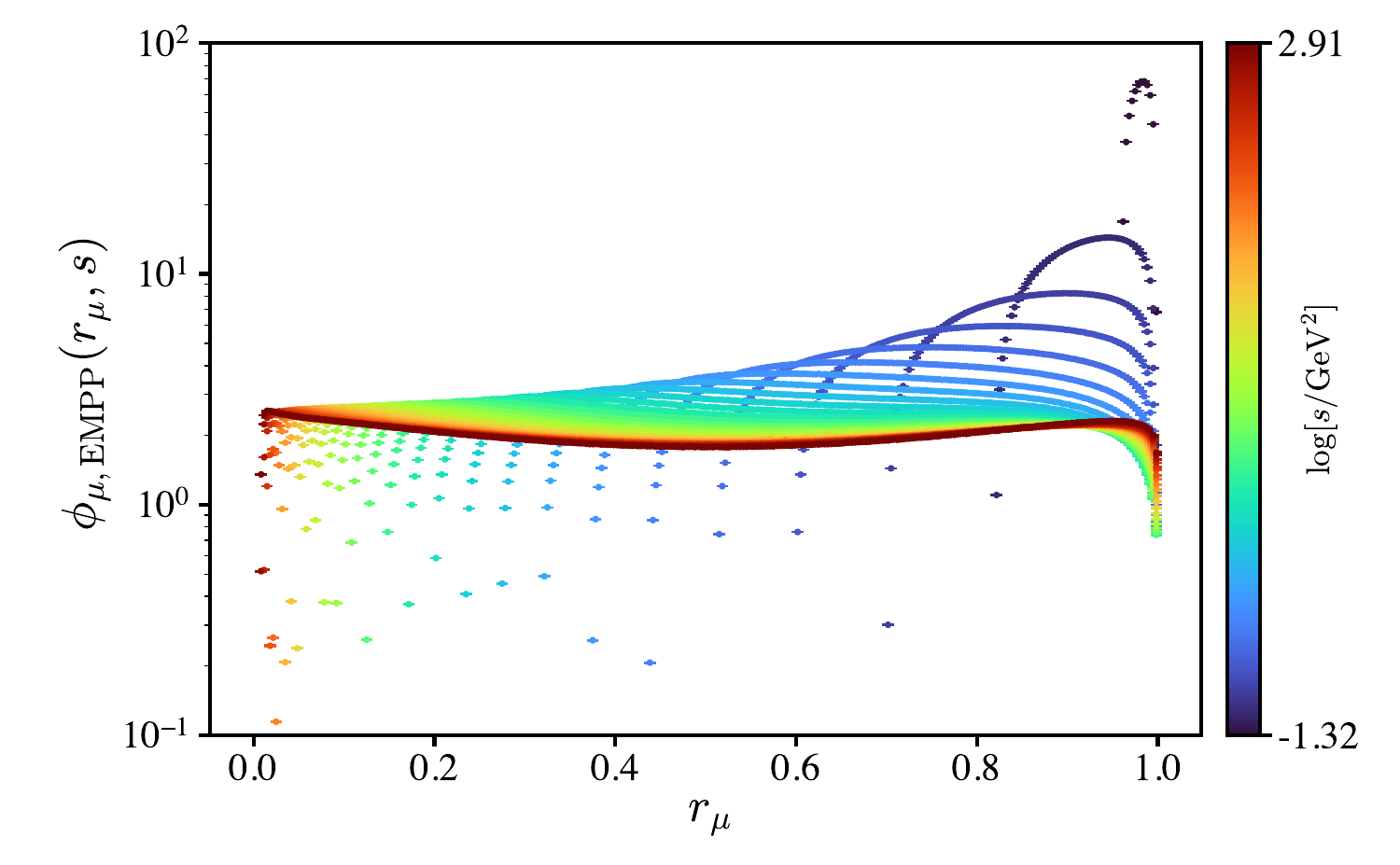}
    \caption{The $\phi_\mu(r_\mu,s)$ for EMPP process, as function of $r_\mu=2E^\ast_\mu/\sqrt{s}$ and for different values of $s$, where $E^\ast_\mu$ is the muon energy in the CoM frame.}
    \label{fig:EMPP_phi_mu}
\end{figure}

\begin{figure}[t!]
    \centering
    \includegraphics[width=0.48\textwidth]{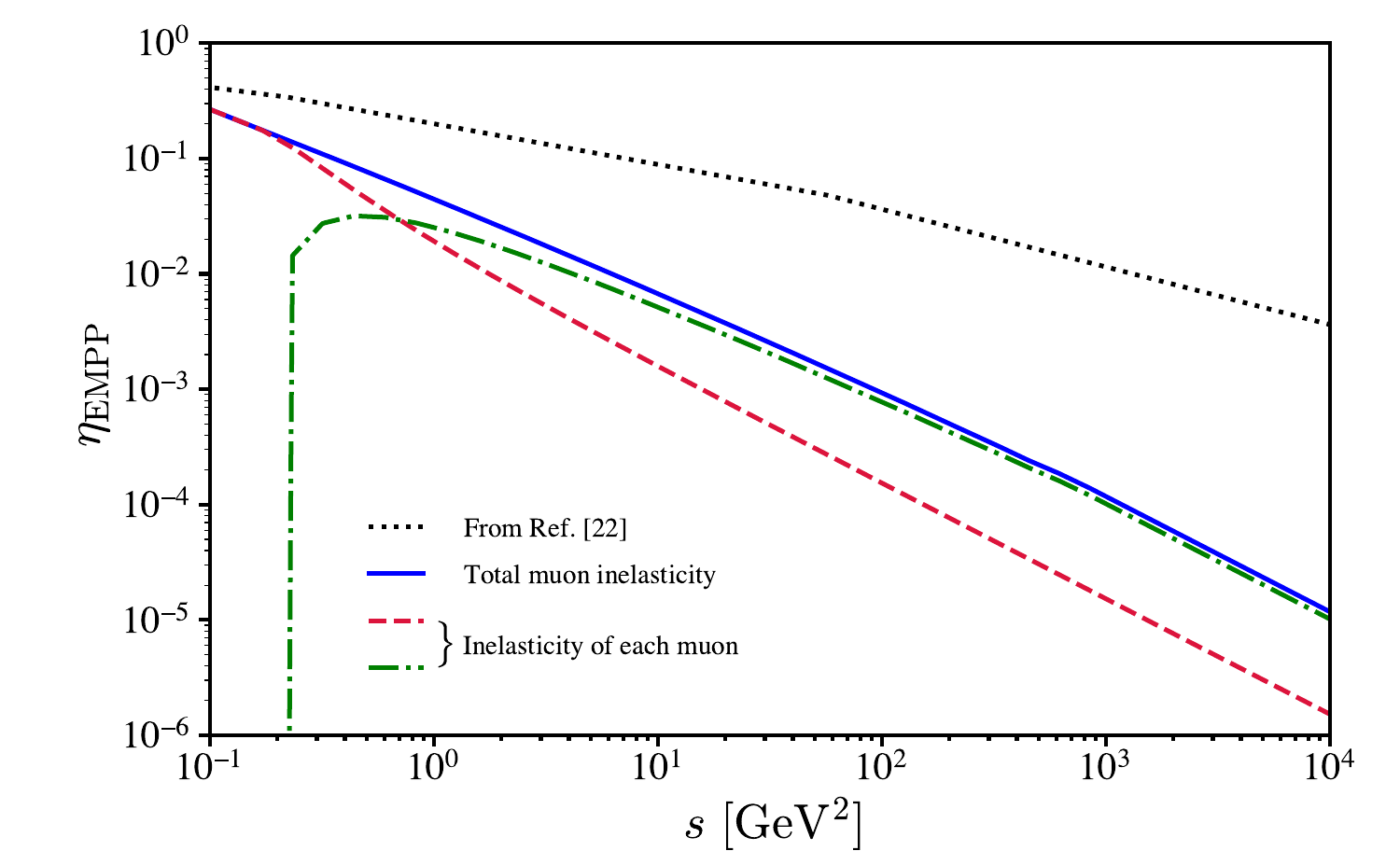}
    \caption{The inelasticities of muons in EMPP. The dashed red and dot-dashed green curves show the inelasticities of the two muons and the sum in shown by the solid blue curve. The dotted black curve depicts the approximation taken from~\cite{Athar:2001sm}.}
    \label{fig:EMPP_eta}
\end{figure}

The functions $\phi_\ell(r_\ell,s)$ for each outgoing particle, in the CoM frame, can be computed numerically. Figure~\ref{fig:EMPP_phi_mu} shows the $\phi_\mu(r_\mu,s)$ as function of $r_\mu$ and for different values of $s$. The energy share of each muon in EMPP can be obtained by changing the integration limit in eq.~(\ref{eq:eta_phi_back}) to $[m_\mu^2/s,0.5]$ ($[0.5,1]$) for the low energy (high energy) muon and multiply it by 2. The inelasticities of the two muons in EMPP and their sum are shown respectively by the dashed red, dot-dashed green and solid blue curves in figure~\ref{fig:EMPP_eta}. The solid blue curve, that is the total inelasticity of the muons, can be compared with the approximation $\eta \approx 3.44\left(s/m_\mu^2\right)^{-0.5}$, at high $s$ values, from~\cite{Athar:2001sm} which is depicted by the dotted black curve in figure~\ref{fig:EMPP_eta}. It turns out that it is acceptable at the $\sim 30\%$ level only at the lowest energies considered here.

\bibliography{references}
\end{document}